# What is the second law of thermodynamics and are there any limits to its validity?

Elias P. Gyftopoulos

*Massachusetts Institute of Technology, Cambridge, MA 02139, USA*

Gian Paolo Beretta

*Università di Brescia, via Branze 38, 25123 Brescia, Italy*

**Abstract**

In the scientific and engineering literature, the second law of thermodynamics is expressed in terms of the behavior of entropy in reversible and irreversible processes. According to the prevailing statistical mechanics interpretation the entropy is viewed as a nonphysical statistical attribute, a measure of either disorder in a system, or lack of information about the system, or erasure of information collected about the system, and a plethora of analytic expressions are proposed for the various measures. In this paper, we present two expositions of thermodynamics (both 'revolutionary' in the sense of Thomas Kuhn with respect to conventional statistical mechanics and traditional expositions of thermodynamics) that apply to all systems (both macroscopic and microscopic, including single particle or single spin systems), and to all states (thermodynamic or stable equilibrium, nonequilibrium, and other states). The first theory is presented without reference to quantum mechanics even though quantum theoretic ideas are lurking behind its definitions. The second theory is a unified quantum theory of mechanics and thermodynamics without statistical probabilities, based on the assumption of a broader set of quantum states than postulated in standard quantum mechanics. We are not presenting another version of statistical quantum mechanics. Here entropy emerges as a microscopic nonstatistical property of matter. Included in the unified theory is an equation of motion that extends the Schrödinger equation of standard quantum mechanics from the domain of zero entropy states to that of nonzero entropy, and depending on initial conditions describes reversible processes that evolve either unitarily or nonunitarily in time, and irreversible processes that evolve nonunitarily in time along the direction of steepest entropy ascent compatible with energy and number of particle conservations.

In the light of these two developments, we discuss some of the comments made in a recent conference sponsored by AIP on "Quantum Limits to the Second Law", and in a book on "Challenges to the Second Law of Thermodynamics", and conclude that there are neither quantum limits nor challenges to thermodynamics, in general, and the second law in particular. The laws of thermodynamics emerge as theorems of our unified, non-statistical quantum theory of mechanics and thermodynamics.

*Key words:* thermodynamics, entropy, second law, quantum mechanics, nonequilibrium, irreversibility

*PACS:* 03.65.-w, 03.65.Ta, 05.45.-a, 05.70.-a

# 1 Introduction

Given that many ideas about thermodynamics were introduced in the 19th and early 20th centuries, we might have assumed that there is little controversy about its foundations and applications. However, even a cursory review of the relevant literature shows that this is not the case. The ideas of thermodynamics have been the subject of controversy ever since their inception [1], controversy that continues even today. Though dated, the following comments continue to be valid. Obert [2] writes: "Most teachers will agree that the subject of engineering thermodynamics is confusing to the student despite the simplicity of the usual undergraduate presentation". Again, Tisza [3] states: "The motivation for choosing a point of departure for a derivation is evidently subject to more ambiguity than the technicalities of the derivation ... In contrast to errors in experimental and mathematical techniques, awkward and incorrect points of departure have a chance to survive for a long time". Werhl [4] writes: "It is paradoxical that although entropy is one of the most important quantities of physics, its main properties are rarely listed in the usual textbooks on statistical mechanics". Lindblad [5] gives a large number of different expressions for entropy and comments: "The entropy function is not unique. Instead there is a family of such functions, one for each set of thermodynamic processes allowed by the experimenter's control of the dynamics of the system through the external fields. This scheme is in line with the philosophy described by Jaynes' dictum: Entropy is a property, not of the physical system, but of the particular experiments you or I choose to perform on it." Truesdell [6] identifies several different statements of the second law. Bunge [7] lists about twenty ostensibly in-equivalent but equally vague formulations of the second law. Mehra and Sudarshan [8] among many other very important recommendations declare: "We maintain that for the explanation of statistical mechanical phenomena the law of evolution is not Hamiltonian, and by creating a generalized dynamics which is essentially non-Hamiltonian we can rid ourselves of all *ad hoc* intermediate assumptions. Thereby we can also shed all the paradoxes that arise in connection with Boltzmann's equation and the *H*-theorem, as well as the pretense of the mechanical explanation of the second law of thermodynamics".

It is noteworthy that the recommendation just cited was made also by Sadi Carnot [9] about a century and a half earlier in his pioneering and trail blazing *"Reflections on the motive power of fire"*. He said: "In order to consider in the most general way the principle of the production of motion by heat, it must be considered independently of any mechanism or any particular agent. It is necessary to establish principles applicable not only to steam engines but to all imaginable heat-engines, whatever the working substance and whatever the method by which it is operated."

And then Carnot continues: "Machines which do not receive their motion from heat ... can be studied even to their smallest details by the mechanical theory. All cases are foreseen, all imaginable movements are referred to these general principles ... . This is the character of a complete theory. A similar theory is evidently needed for heat-engines. We shall have it only when the laws of physics shall be extended enough, generalized enough, to make known before hand all the effects of heat acting in a determined manner on any body".

*Email addresses:* epgyft@aol.com (Elias P. Gyftopoulos), beretta@unibs.it (Gian Paolo Beretta ).

About forty years ago, a group at MIT began a response to the concerns just cited which we describe below not necessarily in the chronological order it was developed. Gyftopoulos and Beretta [10] have composed an exposition of thermodynamics in which all basic concepts are defined completely and without circular and ambiguous arguments in terms of the mechanical ideas of space, time, and force or inertial mass only, Hatsopoulos and Gyftopoulos [11] conceived of a unified quantum theory of mechanics and thermodynamics, and Beretta [12] discovered a complete equation of motion for the unified quantum theory just cited. Both quantum mechanical developments involve only quantum mechanical and no statistical probabilities of the types used in either statistical classical mechanics, or statistical quantum mechanics.

In what follows, we present the foundations of the new theory of thermodynamics without reference to quantum mechanics in Sections 2 and 3, the foundations of the unified quantum theory of mechanics and thermodynamics without the complete equation of motion in Section 4, the complete equation of motion in Section 5, our observations about limits and challenges to the second law in Sections 6 and 7, and our conclusions in Section 8.

## 2 General Foundations of Thermodynamics

### 2.1 General Remarks

In this Section, we outline the exposition of the foundations of thermodynamics discussed in Ref. 10. The order of introduction of concepts and principles is unconventional, designed so as to avoid the typical logical loops and ambiguities of the traditional expositions, and to allow a general definition of entropy valid for all states, including nonequilibrium, from which we derive precise definitions of all other concepts, such as temperature, heat interactions, the simple system model for non-reacting pure substances and mixtures, and for reacting mixtures. The order of introduction is: system (types and amounts of constituents, forces between constituents, and external forces or parameters); properties; states; first law (without energy, work, and heat); energy (without work and heat); energy balance; classification of states in terms of time evolutions; stable equilibrium states; second law (without temperature, heat, and entropy); generalized available energy; entropy of any state (stable equilibrium or not) in terms of energy and generalized available energy, and not in terms of temperature and heat; entropy balance; fundamental relation for stable equilibrium states only; temperature, total potentials (chemical and/or electrochemical), and pressure in terms of energy, entropy, amounts of constituents and parameters for stable equilibrium states only; third law; work in terms of energy; and heat in terms of energy, entropy, and temperature, bulk flow states and interactions, the simple system model for systems with large amounts of constituents, and its extension for systems with chemical reactions.

All concepts and principles are valid for all systems (both microscopic and macroscopic), all states (both thermodynamic or stable equilibrium states, and states that are not stable equilibrium).

### 2.2 Definition of thermodynamics

We define *general thermodynamics* or simply *thermodynamics* as the study of physical observables of motions of physical constituents (particles and radiations) resulting from externally applied forces, and from internal forces (actions on and reactions between constituents). This definition is identical to that given by Timoshenko and Young about mechanical dynamics [13]. However, we will see that the definition encompasses a much broader spectrum of phenomena than mechanical dynamics.

## 2.3 Kinematics: conditions at an instant in time

In our exposition, we give precise definitions of the terms system, property, and state so that each definition is valid without change in any paradigm of physics, and involves no statistics attributable to lack of either computational abilities, or lack of information about any aspect of a problem in physics, and/or consideration of numerical and computational difficulties. Our definitions include innovations. To the best of our knowledge, they violate no theoretical principle, and contradict no experimental results.

A *system* is defined as a collection of *constituents*, subject to both *internal forces*, that is, forces between constituents, and *external forces* that may depend on geometrical or other types of parameters but must be independent of coordinates of constituents that do not belong to the collection. Everything that is not included in the system constitutes its *environment*. A very important consequence of the definition of system to be valid is that the system be both separable from and uncorrelated with its environment.

For a system with r constituents, we denote the amounts by the vector $\mathbf{n} = \{n_1, n_2, ..., n_r\}$. For a system subject to external forces described by s parameters we denote the parameters by the vector $\boldsymbol{\beta} = \{\beta_1, \beta_2, ..., \beta_s\}$. One parameter may be volume $\beta_1 = V$ and the geometric features that define the volume, another may be the potential of an externally determined electric, magnetic or gravitational field, $\beta_2 = \phi$.

At any instant in time, the amounts of the constituents and the parameters of each external force have specific values. We denote these values by $\mathbf{n}$ and $\boldsymbol{\beta}$ with or without additional subscripts.

By themselves, the values of the amounts of constituents and of the parameters at an instant in time do not suffice to characterize completely the condition of the system at that time. We also need the values of a complete set of independent properties at the same instant in time. A *property* is defined by means of a set of measurements and operations – a *measurement procedure* – that are performed on the system and result in a numerical value – the *value of the property* – that can be evaluated at any given instant in time (not as an average over time) and that is independent of: (1) the measuring devices used to implement the measurement procedure; (2) the results of any measurement procedure performed on any other system in the environment; and (3) other instants in time.

The instantaneous values of the amounts of the constituents, the values of the parameters, and the values of a complete set of independent properties encompass all that can be said about the system at a given instant in time, that is, about the results of any measurements that may be performed on the system at that instant in time. We call this set of instantaneous values the *state* of the system at the given instant in time, provided the results of measurements on the system are not correlated with measurements on any other systems in its environment. This definition of state, without change, applies to all paradigms of physics.

The conditions of separability from constituents not included in the system, and of absence of correlations between the system and systems in its environment imply very important restrictions. For example, energy cannot be well defined for a non-separable collection of constituents because it does not satisfy the restrictive definition of system. Again, entropy cannot be well defined for a non-uncorrelated collection of constituents because it does not satisfy the restrictive definition of state. Unfortunately, overlooking these restrictions has often been a source of confusion and erroneous conclusions about the implications of thermodynamics.

## 2.4 Dynamics: changes of state in time

The state of a system may change in time either spontaneously due to the internal forces or as a result of interactions with other systems, or both.

The relation that describes the evolution of the state of either an isolated system – *spontaneous changes of state* – or of a system subject to forces that do not violate the definition of the system is the *equation of motion*. Certain time evolutions obey Newton's equation which relates the force $F$ on each system particle to its mass $m$ and acceleration $a$ so that $F = ma$. Other evolutions obey the Schrödinger equation, that is, the quantum-mechanical equivalent of Newton's equation or Hamilton's equations. Both equations describe processes that are reversible, and more specifically the Schrödinger equation describes reversible processes that evolve unitarily in time. But not all reversible processes evolve unitarily, and not all processes are reversible. So a more complete equation is needed. A response to this need is discussed later.

For the nonce, we note that many features of the complete equation of motion have already been discovered. These features provide not only guidance for the discovery of the complete equation but also a powerful alternative procedure for analyses of many time dependent, practical problems. Two of the most general and well-established features are captured by the three laws of thermodynamics discussed later and their consequences.

## 2.5 Energy and energy balance

Energy is a concept that underlies our understanding of all physical phenomena, yet its meaning is subtle and difficult to grasp. It emerges from a fundamental principle known as the first law of thermodynamics.

The *first law* asserts that any two states of a system may always be the initial and final states of a weight process. Such a process involves no net effects external to the system except the change in elevation between $z_1$ and $z_2$ of a weight, that is, solely a mechanical effect. Moreover, for a given weight, the value of the expression $Mg(z_1 - z_2)$ is fixed only by the end states of the system, where $M$ is the mass of the weight, and $g$ the gravitational acceleration. Many other mechanical effects can be used in the statement of the first law instead of the weight.

The main consequence of this law is that every well-defined system $A$ in any well-defined state $A_1$ has a property called *energy*, with a value denoted by the symbol $E_1$. The energy $E_1$ can be evaluated by a weight process that connects $A_1$ and a reference state $A_0$ to which is assigned an arbitrary reference value $E_0$ so that

$$E_1 = E_0 - Mg(z_1 - z_0) \tag{1}$$

Energy is shown to be an *additive property* [10], that is, the energy of a composite system is the sum of the energies of the subsystems. Moreover, it is also shown that energy has the same value at the final time as at the initial time if the system experiences a zero-net-effect weight process, and that energy remains invariant in time if the process is spontaneous. In either of the last two processes, $z_2 = z_1$ and $E(t_2) = E(t_1)$ for time $t_2$ greater than $t_1$, that is, energy is *conserved*. Energy conservation is a time-dependent result. In Ref. 10, this result is obtained without use of the complete equation of motion.

Energy is transferred between systems as a result of interactions. Denoting by $E^{A\leftarrow}$ the amount of energy transferred from the environment to system $A$ in a process that changes the state of $A$ from $A_1$ to $A_2$, we can derive the *energy balance*. This derivation is based on the additivity of energy and energy conservation, and reads

$$(E_2 - E_1)_{\text{system } A} = E^{A\leftarrow} \tag{2}$$

In words, the energy change of a system must be accounted for by the energy transferred across the boundary of the system, and the arrow indicates that $E^{A\leftarrow}$ is positive if energy is transferred into the system.

*2.6 Types of states*

Because the number of independent properties of a system may be infinite even for a system consisting of a single particle with a single translational degree of freedom – a single dimension in which the particle is allowed to move – and because most properties can vary over a range of values, the number of possible states of a system is infinite. The discussion of these states is facilitated if they are classified into different categories according to evolutions in time. This classification brings forth many important aspects of physics, and provides a readily understandable motivation for the introduction of the second law of thermodynamics.

The classification consists of unsteady states, steady states, nonequilibrium states, and equilibrium states. *Unsteady* and *steady states* occur as a result of sustained (continuous) interactions of the system with other systems in the environment. A *nonequilibrium state* is one that changes spontaneously in time, that is, a state that evolves in time without any effects on or interactions with any systems in the environment. An *equilibrium state* is one that does not change in time while the system is isolated – a state that does not change spontaneously. An *unstable equilibrium state* is an equilibrium state that may be caused to proceed spontaneously to a sequence of entirely different states by means of a minute and short-lived interaction that has either an infinitesimal and temporary effect or a zero net effect on the state of the environment. A *stable equilibrium* state is an equilibrium state that can be altered to a different state only by interactions that leave net effects in the environment of the system. These definitions are identical to the corresponding definitions in mechanics but include a much broader spectrum of states than those encountered in mechanics. The broader spectrum is due to both the first law and the second law discussed later.

Starting either from a nonequilibrium state or from an equilibrium state that is not stable, experience shows that energy can be transferred out of the system and affect a mechanical effect without leaving any other net changes in the state of the environment. In contrast, starting from a stable equilibrium state, experience shows that a system cannot affect the mechanical effect just cited. This impossibility is one of the most striking consequences of the first and second laws of thermodynamics.

*2.7 The second law and generalized available energy*

The existence of stable equilibrium states is not self-evident. It was recognized by Hatsopoulos and Keenan [14] as the essence of the second law. Gyftopoulos and Beretta [10] concur with this recognition, and state the *second law* as follows (simplified version): Among all the states of a system with given values of energy, the amounts of constituents, and the parameters, there exists one and only one stable equilibrium state. For each set of the conditions just cited, the stability is not local but global [15, 16]

The existence of stable equilibrium states for the conditions specified and therefore the second law cannot be derived from the laws of mechanics. Within mechanics, the stability analysis yields that among all the allowed states of a system with fixed values of amounts of constituents and parameters, the only stable equilibrium state is that of lowest energy. In contrast, the second law avers the existence of a globally stable equilibrium state for each value of the energy. As a result, for every system the second law implies the existence of a broad class of states in addition to the states contemplated by mechanics.

The existence of stable equilibrium states for various conditions of a system has many theoretical and practical consequences. One consequence is that, starting from a stable equilibrium state of any system, no energy is available to affect a mechanical effect while the values of the amounts of constituents, the internal forces and the parameters of the system experience no net changes [10]. This consequence is often referred to as the impossibility of the perpetual motion machine of the second kind (PMM2). In many expositions of thermodynamics, it is erroneously taken as the statement of the second law. In this exposition, it is only a theorem of our statements of both the first and the second laws [1] and the set of rigorous definitions that provide their rigorous framework. Moreover, it does not suffer the circularity inherent in the Kelvin-Planck statement of the second law.

Another consequence is that not all states of a system can be changed to a state of lower energy by means of a mechanical effect. This is a generalization of the impossibility of a PMM2. In essence, it is shown that a novel important property exists which is called *generalized adiabatic availability* and denoted by $\Psi$ [10]. The generalized adiabatic availability of a system in a given state represents the optimum amount of energy that can be exchanged between the system and a weight in a weight process that begins with system $A$ in a state $A_1'$ with values $\mathbf{n}_1'$, $\boldsymbol{\beta}_1'$, and ends in a state $A_2''$ with values $\mathbf{n}_2''$, $\boldsymbol{\beta}_2''$. Like energy, this property is well defined for all systems and all states, but unlike energy it is not additive [10].

In striving to define an additive property that captures the important features of generalized adiabatic availability, Gyftopoulos and Beretta introduce a special reference system, called a *reservoir*, and discuss the possible weight processes that the composite of a system and a reservoir may experience. Thus they disclose another consequence of the first and second laws, that is, a limit on the optimum amount of energy that can be exchanged between a weight and a composite of a system and a reservoir [2] $R$ – the optimum mechanical effect. They call the optimum value *generalized available energy*, denote it by $\Omega^R$, and show that it is additive and a generalization of the concept of motive power of fire introduced by Carnot. It is a generalization because he assumed that both systems of the composite acted as reservoirs with fixed values of their respective amounts of constituents and parameters, whereas Gyftopoulos and Beretta do not use this assumption.

The detailed definition of a reservoir is discussed in Ref. 10.

For an *adiabatic process* [3] of system $A$, it is shown that the energy change $E_2 - E_1$ of $A$, the adiabatic availability change $\Psi_2 - \Psi_1$ of $A$, and the generalized available energy change $\Omega_2^R - \Omega_1^R$ of the composite of $A$ and reservoir $R$ satisfy the following relations. If the adiabatic process of $A$ from state $A_1$ to state $A_2$ is reversible,

$$E_2 - E_1 = \Psi_2 - \Psi_1 = \Omega_2^R - \Omega_1^R \tag{3}$$

If instead the adiabatic process of $A$ is irreversible, then

$$E_2 - E_1 > \Psi_2 - \Psi_1 \qquad \text{and} \qquad E_2 - E_1 > \Omega_2^R - \Omega_1^R \tag{4}$$

A process is *reversible* if both the system and its environment can be restored to their respective initial states. A process is *irreversible* if the restoration just cited is impossible. In either process, the restoration path need not and usually does not coincide with the reverse of the initial sequence of states. Relations (3) and (4) provide quantitative criteria to ascertain whether a given adiabatic process for $A$ from $A_1$ to $A_2$ is reversible or irreversible.

---

[1] Incidentally, this conclusion is all that is needed to exorcise Maxwell's demon on the basis of non-quantum mechanical arguments.

[2] A *reservoir* is a special limiting system such that: (1) whatever interaction it experiences it passes through a sequence of stable equilibrium states; and (2) identical reservoirs are always in mutual equilibrium even if in different states.

[3] A process is *adiabatic* if its effects (changes of state of the system and its environment) can be reproduced by a sequence of: (i) a weight process for the system; and (ii) a weight process for its environment.

It is noteworthy that energy, adiabatic availability, and generalized available energy are defined for any system in any state, regardless of whether the state is steady, unsteady, nonequilibrium, equilibrium, or stable equilibrium, and regardless of whether the system has many degrees of freedom or one degree of freedom, or whether the number of constituents and/or the volume of the system are large or small.

A further important result is the definition of a property, that we denote by $c_R$, defined for all states of a reservoir $R$ by a measurement procedure that compares its energy change $(\Delta E_{12}^R)_{\text{rev}}^A$ in a reversible weight process for the composite of $R$ and an auxiliary system $A$ in which $A$ changes from fixed states $A_1$ and $A_2$ with the energy change $(\Delta E_{12}^{R_o})_{\text{rev}}^A$ of a reference reservoir $R_o$ under otherwise identical conditions. Indeed, the ratio of these two energy changes turns out to assume the same value for all states of a reservoir $R$ (regardless of the choice of the fixed states $A_1$ and $A_2$ of the auxiliary system $A$) and, being dimensionless and hence independent of the dimensions used in non-thermodynamic theories of physics, it requires the introduction of a new fundamental dimension defined by comparison with a selected reference reservoir to which we assign an arbitrary value of $c_{R_o}$, that is, $c_R$ is defined as

$$c_R = c_{R_o} \frac{(\Delta E_{12}^R)_{\text{rev}}^A}{(\Delta E_{12}^{R_o})_{\text{rev}}^A} \tag{5}$$

This property $c_R$ turns out to have a positive constant value for all stable equilibrium states of a given reservoir $R$. Moreover, because water at the triple point approximates the definition of a reservoir – it behaves as a reservoir when finite amounts of the solid, liquid and vapor phases are present – it can be selected as the reference reservoir $R_o$.

Because later on, after we define the concept of temperature for the stable equilibrium states of any system, we prove [10] that the value and the dimension of property $c_R$ are equal to the value and the dimension of the temperature of every stable equilibrium state of a given reservoir $R$, to the triple-point-water reference reservoir $R_o$ we may assign arbitrarily (for obvious historical reasons) the value $c_{R_o}$ = 273.16 K, so as to introduce the *kelvin* as a unit of measure of property $c_R$. This constant property of every reservoir enters the general definition of entropy introduced in the next section.

## 3 Entropy

### 3.1 Definition

A system $A$ in any state $A_1$ has many properties. Two of these properties are energy $E_1$ and generalized adiabatic availability $\Psi_1$. Also a composite of a system $A$ and a reservoir $R$ has many properties. One of these properties is *generalized available energy* $\Omega^R$ with respect to the given reservoir $R$. The two properties $E$ and $\Omega^R$ determine a property of $A$ only, which is called *entropy*, and denoted by the symbol $S$. The entropy $S_1$ of state $A_1$ can be evaluated by means of any reservoir $R$, a reference state $A_0$ to which we assign arbitrary values of energy $E_0$, entropy $S_0$, and generalized available energy $\Omega_0^R$, and the expression:

$$S_1 = S_0 + \frac{1}{c_R}[(E_1 - E_0) - (\Omega_1^R - \Omega_0^R)] \tag{6}$$

where $c_R$ is the constant property of the auxiliary reservoir $R$ defined earlier. Entropy $S$ is shown to be independent of the reservoir, that is, $S$ is an inherent – intrinsic – property of system $A$ only like energy $E$ is an inherent property of system $A$ only. The reservoir has an auxiliary role and is used only because it facilitates the definition of $S$ and the understanding of its very important physical meaning.[4] It is also shown that $S$ can be assigned absolute values that are non-negative, and that vanish for all the states encountered in mechanics, quantum or other.

Because energy and generalized available energy satisfy relations (3) and (4), the entropy defined by Eq. (6) remains invariant in any reversible adiabatic process of $A$, and increases in any irreversible adiabatic process of $A$. These conclusions are valid also for spontaneous processes, and for zero-net-effect interactions. The latter features are known as *the principle of non-decrease of entropy*. Both a spontaneous process and a zero-net-effect interaction are special cases of an adiabatic process of system $A$.

The entropy created as a system proceeds from one state to another during an irreversible process is called *entropy generated by irreversibility*. It is positive. The entropy non-decrease is a time-dependent result. In our exposition of thermodynamics in Ref. 10, this result is obtained without use of the complete equation of motion but as a consequence of the statements of the first and the second laws together with the carefully worded set of noncircular definitions just outlined. Because both energy and generalized available energy are additive, Eq. (6) implies that entropy is also additive.

Like energy, entropy can be transferred between systems by means of interactions. Denoting by $S^{A\leftarrow}$ the amount of entropy transferred from systems in the environment to system $A$ as a result of all interactions involved in a process in which the state of $A$ changes from state $A_1$ to state $A_2$, we derive a very important analytical tool, the *entropy balance*, that is

$$(S_2 - S_1)_{\text{system A}} = S^{A\leftarrow} + S_{\text{irr}} \tag{7}$$

where $S_{\text{irr}}$ is non-negative. A positive $S_{\text{irr}}$ represents the entropy generated spontaneously within system $A$ in the time interval from $t_1$ to $t_2$ required to affect the change from state $A_1$ to state $A_2$. Spontaneous entropy generation within a system occurs if the system is in a nonequilibrium or an unstable equilibrium state and then the internal system dynamics precipitates the natural tendency toward stable equilibrium.

The dimensions of $S$ depend on the dimensions of both energy and $c_R$. From Eq. (6), it follows that the dimensions of $S$ are energy over temperature. Because $E$ and $\Omega^R$ are well defined for all states of all systems, it is clear that Eq. (6) defines the entropy $S$ for all states, steady, unsteady, nonequilibrium, metastable and unstable equilibrium, and stable (thermodynamic) equilibrium.

## 3.2 Stable equilibrium states

Among the many states of a system that have given values of the energy $E$, the amounts of constituents $\mathbf{n}$, and the parameters $\boldsymbol{\beta}$, it is shown that the entropy of the unique stable equilibrium state that corresponds to these values is larger than that of any other state with the same values $E$, $\mathbf{n}$, and $\boldsymbol{\beta}$,[5] and can be expressed as a function

$$S = S(E, \mathbf{n}, \boldsymbol{\beta}) \tag{8}$$

---

[4] Indeed, apart from the constants $S_0$ and $(E_0 - \Omega_0^R)/c_R$ in Eq. (6), entropy emerges as a direct measure of the ratio between the energy of the system that is *not* available with respect to a given reservoir $R$, that is, $E_1 - \Omega_1^R$, and the reservoir constant $c_R$. This ratio is independent of the choice of the reservoir $R$, that is, for the given state $A_1$ of system $A$, $(E_1 - \Omega_1^R)/c_R = (E_1 - \Omega_1^{R'})/c_{R'}$ for any pair of reservoirs $R$ and $R'$.
[5] This result is called the *highest-entropy principle* [10]

Equation (8) is called the *fundamental relation*.

The fundamental relation is a concave analytic function with respect to each of its variables. [6] In particular [10],

$$\left[\frac{\partial^2 S}{\partial E^2}\right]_{\mathbf{n},\boldsymbol{\beta}} \leq 0 \qquad (9)$$

Moreover, the fundamental relation is used to define other properties of stable equilibrium states, such as *temperature T*

$$\frac{1}{T} = \left[\frac{\partial S}{\partial E}\right]_{\mathbf{n},\boldsymbol{\beta}} \qquad (10)$$

*total potentials* $\mu_i$

$$\mu_i = -T\left[\frac{\partial S}{\partial n_i}\right]_{E,\mathbf{n},\boldsymbol{\beta}} \quad \text{for } i = 1, 2, .., r \qquad (11)$$

and *pressure p*

$$p = T\left[\frac{\partial S}{\partial V}\right]_{E,\mathbf{n},\boldsymbol{\beta}} \quad \text{for } \beta_1 = V = \text{(volume)} \qquad (12)$$

Temperature, total potentials, and pressure of a stable equilibrium state appear in the necessary conditions for systems to be in mutual stable equilibrium if they interact by exchanging energy, entropy, amounts of constituents, and volume. The conditions are temperature equality, total potential equalities, and pressure equality. Moreover, these equalities are the bases for the measurements of $T$, $\mu_i$'s, and $p$.

The ranges of values of $T$, $\mu_i$'s, and $p$ are infinite. This fact can be established by careful considerations of the behavior of interactions induced by differences in temperature, total potentials, and pressure, and easily proven by quantum mechanical analyses. If quantum mechanical concepts are not used, then the ultimate values of temperature are expressed in the form of the *third law* of thermodynamics as follows. For each given set of values of the amounts of constituents, the internal forces, and the parameters of a system without upper limit on its energy, there exists one stable equilibrium state with infinite inverse temperature, or, equivalently, zero temperature. For systems with both a lower and an upper limit on energy, such as a system having a finite number of energy levels or a system of a finite number of spins only, then there exist two stable equilibrium states with extreme temperatures, one with a positive inverse temperature equal to $+\infty$, and the other with a negative inverse temperature equal to $-\infty$. For all states of a reservoir it is shown that the constant property $c_R$ introduced earlier is the fixed temperature of the reservoir.

### 3.3 Work and heat

A system can experience a great variety of interactions with systems in its environment. Here, we discuss only two of these interactions, work and heat.

---

[6] The proof of the analyticity of $S = S(E, \mathbf{n}, \boldsymbol{\beta})$ requires consideration of the explicit quantum theoretical expressions of energy, amounts of constituents and entropy developed in Section 4 and 5 in terms of the density operator. Although relatively simple and mathematically elegant, the proof is beyond our scope here.

*Work* is an interaction in which the system exchanges only energy with systems in its environment. Thus, for example exchanges of entropy and amounts of constituents are excluded. If the amount of energy exchanged is $W^{\rightarrow}$, then the energy and entropy balances are

$$E_{\text{final}} - E_{\text{initial}} = \Delta E_{\text{system}} = -W^{\rightarrow} \qquad (13)$$
$$S_{\text{final}} - S_{\text{initial}} = \Delta S_{\text{system}} = S_{\text{irr}} \qquad (14)$$

where the arrow indicates that $W^{\rightarrow}$ is positive if energy flows out of the system, and therefore the energy of the system decreases.

*Heat* is an interaction in which the system exchanges only energy and entropy with a reservoir in its environment and nothing else. The amount of energy exchanged is denoted by $Q^{\leftarrow}$ and the entropy by $Q^{\leftarrow}/T_R$, where $T_R$ is the fixed temperature of the reservoir, and the arrow indicates that $Q^{\leftarrow}$ is positive if energy flows into the system. It is noteworthy that $Q^{\leftarrow}$ is not a function of $T_R$ because, by definition, a reservoir has the same $T_R$ for any value $Q^{\leftarrow}$. Thus, for a system experiencing only a heat interaction, the energy and entropy balances are

$$\Delta E_{\text{system}} = Q^{\leftarrow} \qquad (15)$$
$$\Delta S_{\text{system}} = Q^{\leftarrow}/T_R + S_{\text{irr}} \qquad (16)$$

Detailed discussions of properties of stable equilibrium states, such as the *fundamental relation*, *temperature*, *total potentials*, and *pressure*, and of different interactions, such as *work* and a more general definition of *heat* are given in [10], Chapters 8-12.

*3.4 Comment*

The definition of entropy introduced here differs from and is more general than the entropy presented in practically all textbooks on physics and thermodynamics. Our general definition does not involve the concepts of temperature and heat [7]; it is not restricted to large systems; it applies to macroscopic as well as microscopic systems, including a system with one spin, or a system with one particle with only one (translational) degree of freedom; it is not restricted to stable (thermodynamic) equilibrium states; it is defined for stable equilibrium and all other types of states because energy and generalized available energy are defined for all states; and most certainly, it is not statistical – it is an intrinsic property of matter and, of course, for stable equilibrium states, assumes the values listed in existing tables of measured properties of different substances.

To emphasize the difference and generality of our concept of entropy, we recall contrary statements by Meixner [17]: "A careful study of the thermodynamics of electrical networks has given considerable insight into these problems and also produced a very interesting result: the nonexistence of a unique entropy value in a state which is obtained during an irreversible process, ..., I would say I have done away with entropy"; by Callen [18]: "it must be stressed that we postulate the existence of the entropy only for equilibrium states and that our postulate makes no reference whatsoever to nonequilibrium states"[8]; and by Lieb and Yngvason [19]: "Once again, it is a good idea to try to understand first the meaning of entropy for equilibrium states – the quantity that our textbooks talk about when they draw Carnot cycles. In this article we restrict our attention to just those states". It is noteworthy, that even from their totally unjustified and limited perspective, Lieb and Yngvason introduce 16 axioms in order to explain their second law of thermodynamics [20].

---

[7] It involves the definition of $c_R$ which turns out to be the constant temperature of the reservoir. However, the reservoir is proven to play only an auxiliary role in Eq. (6) [10].
[8] In Callen's terminology, equilibrium state means thermodynamic equilibrium state. The same terminology is used in [19].

## 4 Foundations of the unified quantum theory of mechanics and thermodynamics

*4.1 General remarks*

In this section we present a summary of a nonrelativistic quantum theory that differs from the presentations in practically every textbook on the subject. The key differences are the discoveries that for a broad class of quantum-mechanical problems: (i) the probabilities associated with ensembles of measurement results at an instant in time require a mathematical representation delimited by but more general than a wave function or projector; and (ii) the evolution in time of the new mathematical representation requires a nonlinear equation of motion delimited by but more general than the Schrödinger equation [21].

Our definitions, postulates, and major theorems of quantum theory are based on slightly modified statements made by Park and Margenau [22], and close scrutiny of the implications of these statements.

In response to the first difference, Hatsopoulos and Gyftopoulos [11] observed that there exist two classes of quantum problems. In the first class, the quantum-mechanical probabilities associated with measurement results are fully described by wave functions or projectors, whereas in the second class the probabilities just cited require density operators that involve no statistical averaging over projectors — no mixtures of quantum and statistical probabilities. The same result emerges from the excellent review of the foundations of quantum mechanics by Jauch [23]. In addition, this difference eliminates the *"monstrosity"* of the concept of mixed state that concerned Schrödinger [24] and Park [25].

In response to the second difference, Beretta [12] conceived a nonlinear equation of motion for nonstatistical density operators that consists of a linear part plus a nonlinear part. The form of the equation is such that the linear part: (1) accounts for unitary evolutions of density operators in time; (2) reduces to the Schrödinger equation for wave functions and projectors; and (3) determines the quantum theoretical analytical form of the entropy of thermodynamics defined in Section 4.3. On the other hand, the complete equation describes both reversible and irreversible processes, entails all the laws of thermodynamics as theorems, and implies that nonequilibrium states evolve in time towards stable or unstable equilibrium along paths of steepest entropy ascent.

*4.2 Kinematics: Definitions, postulates, and theorems at an instant in time*

*System*

The term is defined exactly the same way as in Section 2.3.

*System postulate*

To every system there corresponds a complex, complete, inner product space, a Hilbert space $\mathcal{H}$. The Hilbert space of a composite system of two distinguishable and identifiable subsystems 1 and 2, with associated Hilbert spaces $\mathcal{H}^1$ and $\mathcal{H}^2$, respectively, is the direct product space $\mathcal{H}^1 \otimes \mathcal{H}^2$.

*Homogeneous or unambiguous ensemble*

At an instant in time, an ensemble of identical systems is called *homogeneous* or *unambiguous* if and only if upon subdivision into subensembles in any conceivable way that does not perturb any member, each subensemble yields in every respect measurement results – spectra of values and frequency of occurrence of each value within a spectrum – identical to the corresponding results obtained from the ensemble. For example, the spectrum of energy levels and the frequency of occurrence of each energy measurement result obtained from any subensemble are identical to the spectrum of energy levels and the frequency of occurrence of each energy measurement result obtained from an independent ensemble that includes all the subensembles. Other criteria are presented in Refs. 11, 12.

*Preparation*

A *preparation* is a reproducible scheme used to generate one or more homogeneous ensembles for study.

Practically every textbook on quantum mechanics after von Neumann [26] avers that the probabilities associated with measurement results [9] of a system in a state "i" are described by a wave function $\Psi_i$ or, equivalently, a projector $|\Psi_i\rangle\langle\Psi_i| = \rho_i = \rho_i^2$, and that each density operator $\rho > \rho^2$ is a statistical average of projectors, that is, each $\rho$ represents a mixture of quantum mechanical probabilities determined by projectors and nonquantum-mechanical or statistical probabilities that reflect our practical inability to handle, control or model all details of a preparation, to control the interactions of the system with its environment, to control initial conditions, to make complex calculations, etc. Mixtures have been introduced by von Neumann [26] for the purpose of explaining thermodynamic equilibrium phenomena in terms of statistical quantum mechanics (see also Jaynes [27] and Katz [28]).

Pictorially, we can visualize a projector by an ensemble of identical systems, identically prepared. Each member of such an ensemble is characterized by the same projector $\rho_i$, and von Neumann calls the ensemble *homogeneous*. Similarly, we can visualize a density operator $\rho$ consisting of a statistical mixture of two projectors $\rho = \alpha_1\rho_1 + \alpha_2\rho_2, \alpha_1 + \alpha_2 = 1, \rho_1 \neq \rho_2 \neq \rho$, where $\rho_1$ and $\rho_2$ represent quantum-mechanical probabilities, $\alpha_1$ and $\alpha_2$ represent statistical probabilities, and the ensemble is called *heterogeneous* or *ambiguous* [11]. A pictorial representation of a heterogeneous ensemble is shown in Figure 1.

These results beg the questions: (i) Are there quantum-mechanical problems that involve probability distributions which cannot be described by a projector but require a purely quantum-mechanical density operator — a density operator which is not a statistical mixture of projectors?; and (ii) Are such purely quantum-mechanical density operators consistent with the foundations of quantum mechanics?

Upon close scrutiny of the definitions, postulates, and key theorems of quantum theory, we find that the answers to both questions are yes. These answers were discovered by Hatsopoulos and Gyftopoulos [11] in the course of their development of a unified quantum theory of mechanics and thermodynamics, and by Jauch [23] in his systematic and rigorous analysis of the foundations of quantum mechanics.

Pictorially, we can visualize a purely quantum-mechanical density operator $\rho > \rho^2$ by an ensemble of identical systems, identically prepared. Each member of such an ensemble is characterized by the same density operator $\rho$ as shown in Figure 2; and by analogy to the results for a projector, we call this ensemble *homogeneous* or *unambiguous* [11]. If the density operator is a projector $\rho_i = \rho_i^2$, then each member of the ensemble is characterized by the same $\rho_i$ as originally proposed by von Neumann.

The recognition of the existence of density operators that correspond to homogeneous ensembles has many interesting implications. It extends quantum ideas to thermodynamics, and

---

[9] We use the expression "probabilities associated with measurement results" rather than "state" because the definition of state requires more than the specification of a projector or a density operator.

# HETEROGENEOUS ENSEMBLE

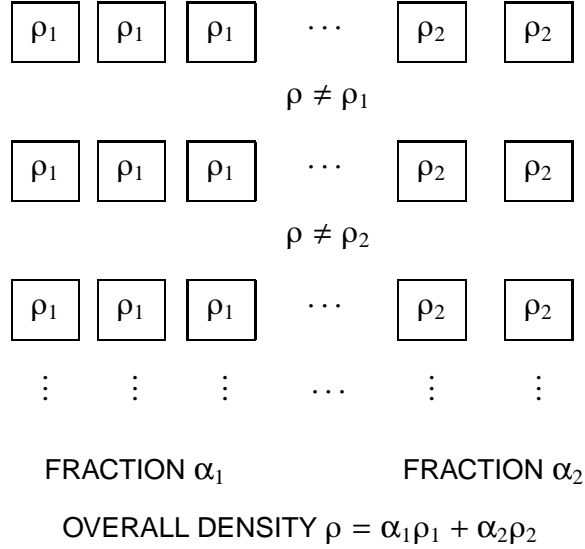

FRACTION $\alpha_1$     FRACTION $\alpha_2$

OVERALL DENSITY $\rho = \alpha_1\rho_1 + \alpha_2\rho_2$

Fig. 1. Pictorial representation of a heterogeneous ensemble. Each of the subensembles for $\rho_1$ and $\rho_2$ represents either a projector ($\rho_i = \rho_i^2$) or a density operator ($\rho_i > \rho_i^2$), for i = 1,2, and $\alpha_1 + \alpha_2 = 1$

# HOMOGENEOUS ENSEMBLE

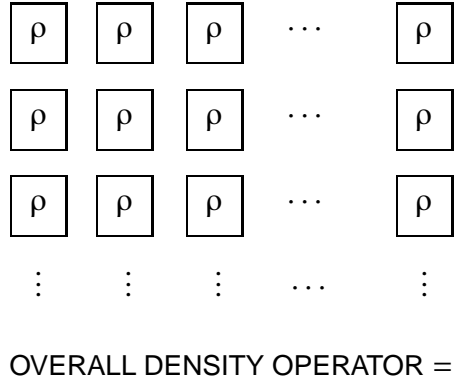

OVERALL DENSITY OPERATOR = $\rho$

Fig. 2. Pictorial representation of a homogeneous ensemble. Each of the members of the ensemble is characterized by the same density operator $\rho \geq \rho^2$. It is clear that any conceivable subensemble of a homogeneous ensemble is characterized by the same $\rho$ as the ensemble.

thermodynamic principles to quantum phenomena. For example, it confirms that thermodynamics applies to all systems (both large and small, including one-constituent systems, such as one spin), to all states (thermodynamic equilibrium, unstable equilibrium, nonequilibrium, steady and unsteady states), and that entropy is an intrinsic property of each constituent of a system [10] (in the same sense that inertial mass is an intrinsic property of each constituent) and not a measure of ignorance, or lack of information, or disorder [29]. In addition and more importantly, it shows that entropy is a measure of the quantum-mechanical spatial shape of the constituents of a system [30–32], and reflects the degree of "involvement" of the constituent particles and their translational and internal-structure degrees of freedom in "load sharing" (equipartition) of the energy and amounts of constituents [33, 34]. Irreversibility is solely due to the changes of this shape as the constituents try to conform to the external and internal force fields of the system [30–32], or in other words, to the spontaneous tendency to improve load sharing among the constituent particles and the accessible levels of their translational and structural modes until their "involvement" is maximally spread compatibly with the internal constraints of the system [33, 34].

*Property*

The term *property* is defined exactly as in Section 2.3. However, perhaps the most striking and distinguished feature of quantum theory is that it brings out explicitly the intrinsic and irreducible probabilistic nature of measurement results, even when performed on a homogeneous ensemble with error-free measurement apparati. As we further discuss below, the main consequence of this well established fact is that the set of measurements and operations that define the value at an instant in time of some properties (that we call *observables*, see below) requires repetition of single measurement acts on a sufficient number of identical systems identically prepared.

In general, the repetition of the same type of measurement on each member of a homogeneous ensemble at one instant in time yields a spectrum of different outcomes. The *value of the property* that is being measured is the arithmetic mean of such measurement results. The restriction to homogeneous ensembles and, in principle, error-free apparati – added to the requirements of independence of: (i) the measuring devices used; (ii) measurements performed on other systems; and (iii) other instants in time – implies that only the probability of outcome of each of the values in the spectrum can be predicted. In what follows we will see that these results define the values of the properties and the state of the system.

*Correspondence postulate and observables*

Some linear Hermitian operators A, B, ... on Hilbert space $\mathcal{H}$, which have complete orthonormal sets of eigenvectors, correspond to properties of a system. Traditionally, properties in this subset are called (quantum theoretical) *observables*.

The inclusion of the word "some" in the correspondence postulate is very important because it indicates that there exist Hermitian operators that do not represent observables, and properties that cannot be represented by Hermitian operators. The first category accounts for Wick et al. [35] superselection rules, and the second accounts both for compatibility of simultaneous measurements introduced by Park and Margenau [22], and for properties, such as temperature, that are not represented by operators and are defined only for restricted families of states, as well as for properties, such as adiabatic availability and entropy, that require measurements of an entire *quorum* of observables [36], i.e., a complete set of linearly independent values of observables sufficient to determine the density operator $\rho$ (see below).

*Measurement act*

A *measurement act* is a reproducible scheme of measurements and operations on a single member of an ensemble. Regardless of whether the measurement refers to an observable or not, in principle the result of such an act is presumed to be precise, that is, an error and perturbation free number belonging to a precise discrete (or continuum) spectrum of possible values dictated solely by the system: the eigenvalues of the associated linear operator.

If a measurement act of an observable is applied to each and every member of a homogeneous ensemble, the results conform to the following postulate and theorems.

*Mean-value postulate*

If a measurement act of an observable represented by a Hermitian operator A is applied to each and every member of a homogeneous ensemble, there exists a linear functional m(A) of A such that the value of m(A) equals the arithmetic mean of the ensemble of A measurements (mean value of the probability distribution function), that is, assuming for simplicity a discrete spectrum of possible outcomes,

$$m(A) = \langle A \rangle = \sum_i a_i/N \text{ for } N \to \infty \qquad (17)$$

where $a_i$ is the result of a measurement act of A applied to the i-th member of the ensemble, a large number (theoretically infinite) of $a_i$'s have the same numerical value, and both m(A) and $\langle A \rangle$ represent the expectation value of A.

*Mean-value theorem*

Each of the mean-value functionals or expectation values m(A) of a system at an instant in time can also be written as

$$m(A) = \langle A \rangle = \text{Tr}[\rho A] \qquad (18)$$

where ρ is an operator that is proven to be Hermitian, positive, unit trace and, in general, not a projector, that is,

$$\rho^\dagger = \rho; \text{ Tr}\rho = 1; \text{ and } \rho \geq \rho^2 \qquad (19)$$

where $\rho^\dagger$ is the Hermitian conjugate of ρ.

The operator ρ is known as the *density operator* or the *density of measurement results of observables*, and here it can be represented solely by a homogeneous ensemble as shown in Figure 2, that is, each member of the ensemble is characterized by the same ρ as any other member. It is noteworthy that the value $\langle A \rangle$ of an observable A depends exclusively on the Hermitian operator A that represents the observable and on the density operator ρ, and not on any other operator that either commutes or does not commute with operator A.

*Probability theorem*

If a measurement act of an observable represented by operator A is applied to each and every member of a homogeneous ensemble characterized by ρ, the probability or frequency of occurrence $W(a_n)$ that the result is $a_n$ is zero unless $a_n$ is an eigenvalue of operator A, in which case it is given by the relation

$$W(a_n) = \text{Tr}[\rho A_n] \qquad (20)$$

where $A_n$ is the projection operator onto the eigenspace $\mathcal{H}_{a_n}$ belonging to $a_n$ and spanned by the set of eigenvectors $|\alpha_n^{(j)}\rangle$ for j=1, 2, ..., $g_n$, where $g_n$ is the degeneracy of $a_n$,

$$A|\alpha_n^{(j)}\rangle = a_n|\alpha_n^{(j)}\rangle \text{ for } n = 1, 2, ... \text{ and } j = 1, 2, ..., g_n \qquad (21)$$

and clearly $A_n = \sum_{j=1}^{g_n} |\alpha_n^{(j)}\rangle\langle\alpha_n^{(j)}|$.

*Measurement result corollary*

The only possible result of a measurement act of the observable represented by A is one of the eigenvalues of A [22].

*State*

The term *state* is defined exactly as in Section 2.3. But here, in view of the correspondence postulate, it consists of the set of values of a quorum of observables, with associated linear Hermitian operators A, B, ..., that is a complete set of linearly independent observables that are sufficient to determine the values of all other observables. For example, for a system with an n dimensional Hilbert space a quorum consists of $n^2-1$ observables. It can be shown that a given set of values of these observables determine uniquely a density operator ρ through the relations

$$\begin{aligned}
\langle A \rangle &= \text{Tr}[\rho A] \\
\langle B \rangle &= \text{Tr}[\rho B] \\
\vdots \quad &\vdots \quad \vdots
\end{aligned} \tag{22}$$

It is noteworthy, that no quantum-theoretic requirement exists which excludes the possibility that the mapping from the measured mean values to ρ must yield a projector $\rho = \rho^2$ rather than a density operator $\rho > \rho^2$, and, conversely, that a pre-specified operator ρ must necessarily be a projector. In fact, the existence of of quantum states and hence homogeneous ensembles that require $\rho > \rho^2$ density operators – that are not derived as a mixture of quantum probabilities and statistical probabilities – provides the means for the unification of quantum theory and thermodynamics without resorting to statistical considerations [11].

*Uncertainty relations*

Ever since the inception of quantum mechanics, the uncertainty relation that corresponds to a pair of observables represented by non-commuting operators is interpreted by many scientists and engineers as a limitation on the accuracy with which the observables can be measured. For example, Heisenberg [37] writes the uncertainty relation between position *x* and momentum $p_x$ in the form

$$\Delta x \, \Delta p_x \geq \hbar/2 \tag{23}$$

and comments: "This uncertainty relation specifies the limits within which the particle picture can be applied. Any use of the words "position" and "velocity" with an accuracy exceeding that given by relation (23) is just as meaningless as the use of words whose sense is not defined."

Again Louisell [38] addresses the issue of uncertainty relations, and concludes: ... "that the actual measurement itself disturbs the system being measured in an uncontrollable way, regardless of the care, skill, or ingenuity of the experimenter. ... In quantum mechanics the precise measurement of both coordinates and momenta is not possible even in principle."

The remarks by Heisenberg and Louisell are representative of the interpretations of uncertainty relations discussed in many textbooks on quantum mechanics [39, 40]. These remarks, however, cannot be deduced from the postulates and theorems of quantum theory.

The probability theorem and the measurement result corollary aver that the measurement of an observable is a precise (perturbation free) and, in many cases, precisely calculable eigenvalue of the operator that represents the observable being measured. So neither a measurement perturbation, a measurement error, and an inaccuracy, nor a practical limitation on the possibility of reproducing identical preparations of identical systems are contemplated by the theorem. An outstanding example of measurement accuracy is the Lamb shift [41].

The probability theorem avers that we cannot predict which precise eigenvalue each single measurement act will yield except in terms of either a prespecified or a measurable probability or frequency of occurrence (or probability distribution function, for observables such as position and momentum that have continuous spectra of possible outcomes). It follows that an ensemble of measurements of an observable, performed on an ensemble of identical systems identically prepared, yields a range of eigenvalues, and a probability or frequency of occurrence distribution over the eigenvalues. In principle, both results are precise and involve no disturbances induced by the measuring procedures. To be sure, each probability distribution of an observable represented by operator X has a variance

$$(\Delta X)^2 = \text{Tr}\rho X^2 - (\text{Tr}\rho X)^2 = \langle X^2 \rangle - \langle X \rangle^2 \tag{24}$$

and a standard deviation $\Delta X$, where $\rho$ is the projector or density operator that describes all the probability distributions of the quantum state in question. Moreover, for two observables represented by two non-commuting operators A and B, that is, such that

$$AB - BA = iC \tag{25}$$

it is readily shown that $\Delta A$ and $\Delta B$ satisfy the uncertainty relation [23, 42, 43]

$$\Delta A\, \Delta B \geq |\langle C \rangle|/2 \tag{26}$$

It is evident that each uncertainty relation refers neither to any errors introduced by the measuring instruments nor to any particular value of a measurement result. The reason for the latter remark is that the value of an observable is determined by the mean value of the operator representing the observable and not by the single outcome of any individual measurement act.

*Collapse of the wave function postulate*

Among the postulates of quantum mechanics, many authoritative textbooks include von Neumann's projection or collapse of the wave function postulate [39, 44]. An excellent, rigorous, and complete discussion of the falsity of the projection postulate is given by Park [45].

*4.3 Dynamics: Linear dynamical postulate for unitary adiabatic processes*

Hatsopoulos and Gyftopoulos [11] postulate that unitary transformations of $\rho$ in time obey the relation

$$\frac{d\rho}{dt} = -\frac{i}{\hbar}[H\rho - \rho H] \tag{27}$$

where H is the Hamiltonian operator of the system. If H is independent of $t$, the unitary evolution of $\rho$ satisfies the equation

$$\rho(t) = U(t, t_0)\rho(t_0)U^\dagger(t, t_0) \tag{28}$$

where $U^\dagger$ is the Hermitian conjugate of U

$$U(t, t_0) = \exp[-(i/\hbar)(t - t_0)H] \tag{29}$$

and if H is explicitly dependent on $t$

$$\frac{dU(t, t_0)}{dt} = -(i/\hbar)H(t)U(t, t_0). \tag{30}$$

Though Eq. (27) looks like the well known von Neumann equation of statistical quantum mechanics, here the equation must be postulated for three reasons: (i) In statistical quantum mechanics [46], the equation is derived as a weighted statistical average of the Schrödinger equation repeatedly applied to describe the evolution of each projector $\rho_i$ representing the components of a heterogeneous ensemble, each multiplied by a time independent statistical weight $\alpha_i$. Here instead $\rho$ is not a mixture of projectors and represents a homogeneous ensemble, therefore it cannot be derived as a weighted statistical average of projectors; (ii) In establishing the von Neumann equation, the statistical probabilities $\alpha_i$ are treated as time independent so that $\sum_i \alpha_i d\rho_i/dt$ can be written as $d\rho/dt$, but then the time evolution is evaluated as if the $\alpha_i$'s were time dependent; and (iii) In postulating Eq. (26), Hatsopoulos and Gyftopoulos recognize that it is limited and incomplete because it describes only adiabatic processes that evolve unitarily in time, and are reversible, and emphasize that not all reversible processes evolve unitarily in time, and not all processes are reversible. So a more general equation of motion is necessary, and is discussed in Section 5.

*4.4 Quantum expression for entropy*

*General criteria for the expression of the entropy*
From the general foundations of thermodynamics outlined and discussed in Sections 2 and 3, we conclude that any expression that purports to represent the entropy $S$ of thermodynamics must have at least the following eight characteristics or, equivalently, conform to the following eight criteria [47].
 (1) The expression must be well defined for every system (large or small) and every state (stable equilibrium and all other states).
 (2) The expression must be invariant in all reversible adiabatic processes, and increase in any irreversible adiabatic process.
 (3) The expression must be additive for all well-defined systems and all well-defined states of such systems (recall that here a system is well defined if it is separable and its state is well defined if it is separable from and uncorrelated with its environment).
 (4) The expression must be non-negative in general, and vanish for all the states encountered in mechanics.
 (5) For given values of the energy, the amounts of constituents, and the parameters, one and only one state must correspond to the largest value of the expression.
 (6) For given values of the amounts of constituents and parameters, the graph of entropy versus energy of stable equilibrium states must be concave and smooth.
 (7) For a composite $C$ of two subsystems $A$ and $B$, the expression must be such that the entropy maximization procedure for $C$ [criterion no. (5)] yields identical thermodynamic potentials (for example, temperature, total potentials, and pressure) for all three systems $A$, $B$, and $C$.
 (8) For stable equilibrium states, the expression must reduce to relations that have been established experimentally and that express the entropy in terms of values of energy, amounts of constituents, and parameters, such as the relations for ideal gases.

It is noteworthy that, except for criteria (1) and (4), we can establish the remaining six criteria by reviewing the behavior of entropy of classical thermodynamics.

*Comment*
It is noteworthy that for a system with given values of energy, volume, and amounts of constituents, if $\rho$ is a projector then the entropy $S=0$, if $\rho$ is neither a projector nor a statistical average of projectors, and corresponds to a state which is not stable equilibrium (not thermodynamic equilibrium), then $S$ has a positive value smaller than the largest possible value for the given specifications, and if $\rho$ corresponds to the unique stable equilibrium state, then $S$ has the largest value of all the entropies of the system which share the given values of energy, volume, and amounts of constituents. Said differently, all projectors or wave functions correspond to zero-entropy physics, all largest-entropy density operators for given system specifications correspond to stable equilibrium states – classical thermodynamics – and for the given specifications, all other density operators that are associated neither with zero entropy, nor with largest-value entropy correspond to probability distributions that can be represented neither by projectors nor by stable-equilibrium-state density operators. The few results just cited indicate that the restriction of quantum mechanics to problems that require probability distributions described only by projectors is both unwarranted and nonproductive.

*Unsatisfactory expressions for the entropy*

Ever since the enunciation of the first and second laws of thermodynamics by Clausius about 140 years ago, all expressions for entropy that are not based on temperature and heat involve probabilities. Invariably, the probabilities are statistical (as opposed to inherent to the nature of physical phenomena), and are justified as a means to partially overcome the enormous computational and informational difficulties resulting from the complexity of large systems. Thus, each expression of entropy is usually construed as a subjective measure of information rather than an analytical description of an intrinsic property of matter.

As we discuss in Section 3, entropy is shown to be an inherent – intrinsic – property of any system (large or small), in any state (stable equilibrium or not), and we aver here that unitary evolutions of $\rho$ in time obey Eq. (27). But in the course of unitary evolutions, the processes are adiabatic and reversible.

In the light of the conclusion just cited, and the eight general characteristics of entropy just outlined, Gyftopoulos and Cubukcu [48] concluded that: (a) Expressions for entropy based on temperature and heat are not acceptable because they are restricted to thermodynamic or stable equilibrium states only; (b) Expressions for entropy proposed in statistical classical mechanics are not acceptable because they are based on statistical (subjective) rather than quantal (inherent) probabilities, and the resulting entropy is not a property of matter; and (c) Expressions for entropy proposed in statistical quantum mechanics [46, 49] that depend on variables other than the eigenvalues of $\rho$ are not acceptable because they fail criterion (2) for adiabatic unitary processes in which the eigenvalues of $\rho$ are time invariant. Accordingly, only some quantum functionals of the density operator $\rho$ that are proposed in the literature qualify as candidates for possible expressions for entropy. They are the following ($k$ is the Boltzmann constant).

The Daróczy entropy [50]

$$S_D = \frac{k}{2^{1-\alpha} - 1}(\text{Tr}\rho^{\alpha-1}) \tag{31}$$

where $\alpha > 0$, $\alpha \neq 1$.

The Hartley entropy [51]

$$S_H = k \ln N(\rho) \tag{32}$$

where $N(\rho)$ is the number of nonzero eigenvalues of $\rho$.

The infinite norm entropy

$$S_\infty = -k \ln \| \rho \|_\infty \tag{33}$$

where $\| \rho \|_\infty$ is the largest eigenvalue of $\rho$.

The Renyi entropy [52]

$$S_R = \frac{k}{1 - \alpha} \ln(\text{Tr}\rho^\alpha) \tag{34}$$

where $\alpha > 0$, $\alpha \neq 1$.

Upon investigating the features of these candidate entropies Gyftopoulos and Cubukcu find that none satisfies all the criteria listed above and therefore they cannot be assumed as representative of the entropy of thermodynamics.

*The unique fully satisfactory expression of the entropy of thermodynamics in terms of the density operator*

As demonstrated by Gyftopoulos and Cubukcu the only functional of the density operator that satisfies all the listed necessary criteria to qualify for being representative of the entropy of thermodynamics is the von Neumann entropy functional [53]

$$S(\rho) = -k\,\mathrm{Tr}\rho\ln\rho \qquad (35)$$

provided that the density operator $\rho$ is not a statistical average of projectors or other density operators, i.e., $\rho$ represents measurements on a homogeneous ensemble.

*Heterogeneous ensembles*

The problem of describing unambiguously measurement results from a heterogeneous ensemble resulting from the statistical mixing of homogeneous ensembles each of which requires a density operator that in general is not a projector, is addressed by Beretta in Ref. 12 where he proposes a measure theoretic mathematical representation that is capable to maintain the quantal probabilities brought in by each component homogeneous ensemble well distinguished from the nonquantal (statistical) mixing probabilities.

## 5 Complete equation of motion

*5.1 General Remarks*

In his doctoral dissertation [12] and subsequent publications [33, 34, 54–59], Beretta addressed successfully the problem of finding the complete equation of motion of quantum thermodynamics, that is, the equation that accounts both for reversible and irreversible evolutions in time of a system that obeys the rules of the nonstatistical unified quantum theory of mechanics and thermodyanmics summarized in Section 4. Prior to discussing the form of the Beretta equation, it is helpful to list the restrictive consistency conditions that must be satisfied by such an equation.

*5.2 A restrictive set of consistency conditions*

In order for an equation of motion of quantum thermodynamics to be acceptable the following conditions must be satisfied [11, 21, 60]:
 (1) If the system experiences no interactions with its environment (and chemical or nuclear reactions are inhibited), energy and amounts of constituents must be conserved.
 (2) If the system is free, momentum conservation and Galilean invariance must be satisfied.
 (3) If the density operator is a projector, that is, $\rho^2 = \rho$, the evolution of $\rho$ in time obeys Eq. 27, and the entropy is zero. This condition preserves all the remarkable successes of standard quantum mechanics, is consistent with experimental results, and (for projector density operators) rules out deviations from linear and unitary dynamics.
 (4) For isolated systems, the rate of change of the entropy functional $S(\rho) = -k\,\mathrm{Tr}\rho\ln\rho$ must be nonnegative.
 (5) For a system with fixed values of energy, amounts of constituents, and parameters, there must be one and only one equilibrium state for which both $d\rho/dt = 0$ and the value of the entropy is larger than that of all the other states with the same values of energy, amounts of constituents, and parameters.

(6) The state of highest entropy just cited must be globally stable with respect to perturbations that do not alter the energy, the amounts of constituents and the parameters [16, 61].

(7) Validity of Onsager reciprocity relations at least for nonequilibrium states in the vicinity of the highest entropy stable equilibrium states.

(8) Existence and uniqueness of well-behaved solutions ρ(*t*) which must remain Hermitian, nonnegative and unit trace for arbitrary initial conditions, that is, for any $\rho(t_0)^\dagger = \rho(t_0) \geq \rho^2(t_0)$ and Trρ($t_0$)=1, and for times smaller and larger than any value $t_0$.

Further conditions are given in Ref. 60, including "conservation of effective Hilbert space dimensionality" that we clarify in Section 5.3, and the conditions for separability and locality that we list in Section 5.5.

As shown in [58] the preceeding conditions are satisfied by the ansatz that the density operator evolves along a trajectory that results from the competition and coexistence of two orthogonal "forces", a Hamiltonian force that tends to drive the density operator along a unitary isoentropic evolution in time, and maintains constant each eigenvalue of ρ, and a conservative but dissipative force that pulls ρ towards the path of steepest entropy ascent.

*5.3 Steepest entropy ascent for a single isolated constituent*

Here we consider only the simplest form of the general equation of motion proposed for the unified quantum theory. For convenience, we define the dimensionless entropy operator S̃=−B lnρ, where B is the idempotent operator obtained from ρ by substituting in its spectral expansion each nonzero eigenvalue with unity. Thus, S̃ is the null operator if $\rho^2=\rho$, and, in general, the entropy functional $-k\,\mathrm{Tr}\rho\ln\rho$ can be written as $-k\,\mathrm{Tr}\rho\tilde{S}$.

For a single isolated constituent without non-Hamiltonian time-invariants, the postulated nonlinear equation of motion coincides with Eq. 27 only for zero entropy states ($\rho^2 = \rho$), whereas, for an arbitrary nonzero entropy state ($\rho > \rho^2$), it is given by the relation

$$\frac{d\rho}{dt} = -\frac{i}{\hbar}[H,\rho] - \frac{1}{\tau}D \tag{36}$$

where $\tau$ is a scalar time constant or functional, D is a nonlinear operator function of ρ, S̃, H defined by any of the following equivalent forms

$$D = -\frac{\begin{vmatrix} \frac{1}{2}\{\tilde{S},\rho\} & \rho & \frac{1}{2}\{H,\rho\} \\ \mathrm{Tr}\rho\tilde{S} & 1 & \mathrm{Tr}\rho H \\ \mathrm{Tr}\rho H\tilde{S} & \mathrm{Tr}\rho H & \mathrm{Tr}\rho H^2 \end{vmatrix}}{\begin{vmatrix} 1 & \mathrm{Tr}\rho H \\ \mathrm{Tr}\rho H & \mathrm{Tr}\rho H^2 \end{vmatrix}} = \frac{\begin{vmatrix} \rho\ln\rho & \rho & \frac{1}{2}\{H,\rho\} \\ \mathrm{Tr}\rho\ln\rho & 1 & \mathrm{Tr}\rho H \\ \mathrm{Tr}\rho H\ln\rho & \mathrm{Tr}\rho H & \mathrm{Tr}\rho H^2 \end{vmatrix}}{(\Delta H)^2}$$

$$= -\frac{1}{2}\{\tilde{S},\rho\} + \alpha\rho + \beta\frac{1}{2}\{H,\rho\} \tag{37}$$

where $|\cdot|$ denotes a determinant, the operators S̃ and ρ commute so that [10] $\frac{1}{2}\{\tilde{S},\rho\} = -\rho\ln\rho$, and

---

[10] Nevertheless we introduce this notation because in Section 5.4, where we generalize the equation to

$\alpha$ and $\beta$ are defined by the nonlinear functionals

$$\alpha = \mathrm{Tr}\rho\tilde{S} - \beta\mathrm{Tr}\rho H \tag{38}$$

$$\beta = \frac{\mathrm{Tr}\rho\tilde{S}H - \mathrm{Tr}\rho\tilde{S}\,\mathrm{Tr}\rho H}{\mathrm{Tr}\rho H^2 - (\mathrm{Tr}\rho H)^2} \tag{39}$$

Eq. (36) satisfies all the consistency conditions listed in Section 5.2. For example, it is easy to verify that

$$\frac{d}{dt}\mathrm{Tr}\rho = \mathrm{Tr}\frac{d\rho}{dt} = \mathrm{Tr}D = 0 \tag{40}$$

$$\frac{d}{dt}\mathrm{Tr}\rho H = \mathrm{Tr}\frac{d\rho}{dt}H = \mathrm{Tr}DH = 0 \tag{41}$$

$$\frac{d}{dt}\mathrm{Tr}\rho\tilde{S} = \mathrm{Tr}\frac{d\rho}{dt}\tilde{S} = \mathrm{Tr}D\tilde{S} \geq 0 \tag{42}$$

In addition, it can be shown that if $\rho > \rho^2$ at one time, it remains so at all times, both forward and backward in time. The only equilibrium density operators that are stable according to this dynamics are the highest entropy operators in the one parameter family

$$\rho^{se} = \frac{\exp(-H/kT)}{\mathrm{Tr}\,\exp(-H/kT)} \tag{43}$$

where the parameter $T$ is readily identified with the temperature defined earlier. [11]

Unitary dynamics (Eq. 27) applied to nonzero entropy states ($\rho > \rho^2$) maintains unchanged each of the eigenvalues of $\rho$. Instead, Eq. (36) maintains at zero only the initially zero eigenvalues of $\rho$ and, therefore, conserves the cardinality of the set of zero eigenvalues, dim Ker($\rho$)=const. This important feature implies that if the isolated system is prepared in a state that does not require all the eigenvectors $|\psi_\ell\rangle$ of H so that $\rho(0)|\psi_\ell\rangle = 0$ for some values of $\ell$, then the zero eigenvalues persist at all times, that is, $\rho(t)|\psi_\ell\rangle = 0$. This is the nontrivial condition that we call conservation of effective Hilbert space [12] dimensionality [and can be viewed as an extension of item (iii) of the consistency conditions listed in Section 5.1] and, of course, it is a characteristic feature of all successful models and theories of physics of isolated systems.

The non-Hamiltonian dissipative term pulls the state operator in the direction of the projection of the gradient of the entropy functional onto the (hyper)plane of constant Tr($\rho$) and Tr$\rho$H. Because the system is isolated, the entropy ceases to increase only when the largest entropy value is reached consistent with the specified dimensionality of the Hilbert space. The same would hold for adiabatic processes described by a time dependent H.

---

composite systems, operators $\tilde{S}$ and $\rho$ are replaced by operators $(\tilde{S})^J$ and $\rho_J$ which in general may not commute.

[11] Proofs of these and other intriguing features of Eq. (36) and its more general forms are given in Refs. 12,16,33,34,54–59,62. In particular, the form of the equation can be readily generalized 33,55 to include other *generators of the motion* in addition to the Hamiltonian operator H, such as the number of particles operators $N_i$ for systems that at stable equilibrium are described by the grand canonical density operator

$$\rho^{se} = \frac{\exp[-(H - \sum \mu_i N_i)/kT]}{\mathrm{Tr}\,\exp[-(H - \sum \mu_i N_i)/kT]} \tag{44}$$

[12] By effective Hilbert space we mean the range Ran$\rho$ of the density operator, namely the subspace of $\mathcal{H}$ spanned by the eigenvectors of $\rho$ with nonzero eigenvalues.

As recently shown by Gheorghiu-Svirschevski [62] and Beretta [33], the steepest-entropy-ascent feature is confirmed also by a variational formulation wherein the form of the dissipative term in Eq. (36) is obtained as a result of searching among all possible directions in which operator ρ can change, the direction of maximal entropy generation compatible with the constraints that ρ remains a well-defined operator, and Trρ and TrρH remain time invariant. For the more general form that conserves also other observables in addition to the energy see Refs. 33, 54.

Given any initial density operator, it is possible to solve the equation of motion not only in forward time but also in backward time [34, 57] and reconstruct the entire trajectory in ρ space for $-\infty < t < \infty$, provided of course either the Hamiltonian H is time independent or its dependence on time is well behaved at all times.

In Ref. 63, Eq. 36 is applied to study atomic relaxation in a two-level atom. By modeling the interaction of a single two-level atom with the quantum electromagnetic field that corresponds to driving the two-level atom near resonance by a nearly monochromatic laser beam, it is shown that the nonlinear irreversible atomic relaxation described by Eq. 36 implies corrections to the resonance fluorescence, absorption and stimulated emission line shapes. Such experiments on properly prepared homogeneous ensembles that require $\rho \neq \rho^2$ would provide a means to evaluate the atomic relaxation time constant $\tau$ in Eq. (36).

*5.4 One particle approximation for a Boltzmann gas*

As an illustration of the applications of Eq. (36) we consider an isolated system composed of non-interacting identical particles with single-particle energy eigenvalues $e_i$ for $i = 1, 2, \ldots, N$ where N is finite and the $e_i$'s are repeated in case of degeneracy. As done by Beretta [34], we restrict for simplicity our analysis on the class of dilute-Boltzmann-gas states in which the particles are independently distributed among the N (possibly degenerate) one-particle energy eigenstates. In density operator language, this is tantamount to restricting the analysis on the subset of one-particle density operators ρ that are diagonal in the representation in which also the one-particle Hamiltonian operator H is diagonal ([H,ρ]=0). We denote by $p_i$ the probability of the i-th energy eigenstate, so that the per-particle energy and entropy functionals are given by the relations

$$E = \sum_{i=1}^{N} e_i p_i \qquad S = -k \sum_{i=1}^{N} p_i \ln p_i \qquad \sum_{i=1}^{N} p_i = 1 \qquad (45)$$

The nonlinear equation of motion maintains the initially zero probabilities equal to zero, whereas the rates of change of the nonzero probabilities are given by

$$\frac{dp_j}{dt} = -\frac{1}{\tau} \frac{\begin{vmatrix} p_j \ln p_j & p_j & e_j p_j \\ \sum p_i \ln p_i & 1 & \sum e_i p_i \\ \sum e_i p_i \ln p_i & \sum e_i p_i & \sum e_i^2 p_i \end{vmatrix}}{\begin{vmatrix} 1 & \sum e_i p_i \\ \sum e_i p_i & \sum e_i^2 p_i \end{vmatrix}} \qquad \text{for i, j} = 1, 2, ..., N \qquad (46)$$

The solutions of these equations are well-behaved in the sense that they satisfy both all the conditions listed in Section 5.1. In particular, as exemplified by the numerical simulations in Ref. 34, they exhibit the following general features: (i) They conserve the energy and trace of $\rho$; (ii) They preserve the non-negativity of each $p_i$; (iii) They maintain the rate of entropy generation non-negative; (iv) They maintain the dimensionality of the effective Hilbert space, that is, for a density operator $\rho$ with $[H,\rho]=0$ and eigenvalues $\mathbf{p}$ given by any set of $p_i$'s they maintain invariant the vector $\boldsymbol{\delta}(\mathbf{p})$ of $\delta_i$'s such that, for each i=1,2,...,N, $\delta_i = 1$ if $p_i \neq 0$ and $\delta_i = 0$ if $p_i = 0$; (v) They drive any arbitrary initial density operator $\rho(t_0)$ towards the partially canonical (or canonical, if $\delta_i = 1$ for all energy eigenstates of the Boltzmann gas) equilibrium density operator $\rho(\infty)$ with time independent eigenvalues $\mathbf{p}(\infty)$ in the energy representation given by

$$p_j^{\text{pe}}(E, \boldsymbol{\delta}) = \frac{\delta_j \exp(-\beta^{\text{pe}}(E, \boldsymbol{\delta}) e_j)}{\sum_{i=1}^{N} \delta_i \exp(-\beta^{\text{pe}}(E, \boldsymbol{\delta}) e_i)} \tag{47}$$

where, $\boldsymbol{\delta} = \boldsymbol{\delta}(\mathbf{p}(0))$, the value of $\beta^{\text{pe}}$ is determined by the initial condition $\sum_{i=1}^{N} e_i p_i^{\text{pe}}(E, \boldsymbol{\delta}) = E = E(\mathbf{p}(0))$, and the superscript pe is used to indicate that the system is in an unstable or, so-called, partial equililbrium state.

Among all the equilibrium states just cited there exists one and only one that is stable (se) and corresponds to the largest value of the entropy for the given value of energy $E$, and for which the eigenvalues of the density operator in the energy representation are given by the canonical distribution

$$p_j^{\text{se}}(E) = \frac{\exp(-e_j/kT(E))}{\sum_{i=1}^{N} \exp(-e_i/kT(E))} \tag{48}$$

where $T(E)$ is shown to be equal to the derivative of energy with respect to entropy of stable equilibrium states of the Boltzmann gas at energy $E$. By definition (see Eq. 10) the derivative just cited is the temperature.

For a general nonequilibrium state, the rate of entropy generation may be written as a ratio of Gram determinants in the form

$$\frac{dS}{dt} = \frac{k}{\tau} \frac{\begin{vmatrix} \sum p_i (\ln p_i)^2 & \sum p_i \ln p_i & \sum e_i p_i \ln p_i \\ \sum p_i \ln p_i & 1 & \sum e_i p_i \\ \sum e_i p_i \ln p_i & \sum e_i p_i & \sum e_i^2 p_i \end{vmatrix}}{\begin{vmatrix} 1 & \sum e_i p_i \\ \sum e_i p_i & \sum e_i^2 p_i \end{vmatrix}} \geq 0 \tag{49}$$

where the non-negativity follows from the well-known properties of Gram determinants.

Given any initial density operator, it is possible to solve the equation of motion for all values of time, that is $-\infty < t < \infty$. In the limit $t \to \infty$ the trajectory approaches a largest entropy equilibrium state with a density operator that is canonical over the energy eigenstates initially included in the analysis. An exception to this conclusion is the case of the initial density operator being a projector $\rho = \rho^2$. Then the evolution in time follows the Schrödinger equation, and is unitary and reversible, except if the projector is an energy eigenprojector which is stationary.

## 5.5 Composite system dynamics

As in standard quantum theory, the composition of a system is embedded in the structure of the associated Hilbert space as a direct product of the subspaces associated with the individual elementary constituent subsystems, as well as in the form of the Hamiltonian operator.

For simplicity, we consider here a system composed of two distinguishable and indivisible elementary constituent subsystems. For example, each subsystem may be a different elementary particle or a Fermi-Dirac or Bose-Einstein or Boltzmann field (in which case the corresponding Hilbert space is a Fock space). The subdivision into elementary constituents, each considered as indivisible, is reflected by the structure of the Hilbert space $\mathcal{H}$ as a direct product of subspaces,

$$\mathcal{H} = \mathcal{H}^A \otimes \mathcal{H}^B \tag{50}$$

and is particularly important because it defines the level of description of the system and specifies its elementary structure, together with the Hamiltonian operator

$$H = H_A \otimes I_B + I_A \otimes H_B + V \tag{51}$$

where $H_J$ is the Hamiltonian operator on $\mathcal{H}^J$ associated with subsystem J when isolated, for J=A, B, and V (on $\mathcal{H}$) is the interaction Hamiltonian among the two subsystems.

The specification just cited determines also the structure of the nonlinear dynamical law, which is different depending on whether the system is or is not subdivisible into indivisible subsystems, i.e., whether or not it has an internal structure. The dependence of the structure of the dynamical law on the level of decription of its internal structure in terms of elementary indivisible constituents is an important consequence of having given up linearity [33, 64].

In the simplest case we are considering here, the dissipative term in the equation of motion is a function of two novel important nonlinear local observables that we call "locally perceived overall-system energy" and "locally perceived overall-system entropy", that represent measures of how the overall-system energy and entropy operators, H and $\tilde{S} = -B \ln \rho$, are "felt" locally within the J-th subsystem [33, 55]. They are associated with the following local operators

$$(H)^J = \mathrm{Tr}_{\bar{J}}[(I_J \otimes \rho_{\bar{J}})H] \tag{52}$$

$$(\tilde{S})^J = \mathrm{Tr}_{\bar{J}}[(I_J \otimes \rho_{\bar{J}})\tilde{S}] \tag{53}$$

where J=A, B and $\bar{J}$ = B, A, $\rho_J = \mathrm{Tr}_{\bar{J}}\rho$ and $\rho_{\bar{J}} = \mathrm{Tr}_J \rho$.

Operator $\tilde{S}_J$ may be interpreted as the subsystem entropy operator only if subsystem J is not correlated with the other subsystem, i.e., only if $\rho$ can be written as

$$\rho = \rho_A \otimes \rho_B \tag{54}$$

then the subsystem entropy is defined and given by the nonlinear state functional of the reduced state operator $k\,\mathrm{Tr}_J \rho_J \tilde{S}_J = -k\,\mathrm{Tr}_J \rho_J \ln \rho_J$, and $\tilde{S} = \tilde{S}_A \otimes I_B + I_A \otimes \tilde{S}_B$. If the subsystems are correlated, then no individual entropies can be defined (note the coherence of this statement with our discussion in Section 3); however, the functional $k\,\mathrm{Tr}_J \rho_J (\tilde{S})^J$ is always well-defined and may be interpreted as the subsystem's local perception of the overall-system entropy.

Similarly, energy is defined for subsystem J only if it is not interacting with the other subsystem, i.e., if H can be written as

$$H = H_A \otimes I_B + I_A \otimes H_B \tag{55}$$

Then the energy of J is given by the functional $\mathrm{Tr}_J \rho_J H_J$. The functional $\mathrm{Tr}_J \rho_J (H)^J$ instead is always well-defined, even if the subsystems are interacting, and may be interpreted as the subsystem's local perception of the overall-system energy.

In order for an equation of motion of quantum thermodynamics to be acceptable for both the description of the time evolution of a composite system, and the exclusion of nonlocality paradoxes such as faster-than-light communication, the following conditions must be added to the list of Section 5.2 [60]:

(1) For a system composed of noninteracting subsystems, the energy of each subsystem, $\text{Tr}_J \rho_J H_J$, must be time invariant (separate energy conservation);
(2) For a system composed of subsystems in independent states, that is, such that $\rho = \rho_A \otimes \rho_B$, the entropy of each subsystem J, $k\,\text{Tr}_J \rho_J \tilde{S}_J$, must be nondecreasing in time (separate entropy nondecrease);
(3) Noninteracting subsystems that are initially in correlated states must be unable to influence each other's time evolution as long as they remain noninteracting, even if each of them separately interacts with other systems.

For an isolated composite of r constituents without non-Hamiltonian time-invariants, the postulated nonlinear equation of motion coincides with Eq. 27 only for zero entropy states ($\rho^2 = \rho$), whereas for an arbitrary nonzero entropy state ($\rho^2 < \rho$) the equation of motion is

$$\frac{d\rho}{dt} = -\frac{i}{\hbar}[H, \rho] - \frac{1}{\tau_A} D_A \otimes \rho_B - \frac{1}{\tau_B} \rho_A \otimes D_B \tag{56}$$

where each $D_J$ is the nonlinear operator defined by the relations

$$D_J = -\frac{1}{2}\{(\tilde{S})^J, \rho_J\} - \alpha_J \rho_J + \beta_J \frac{1}{2}\{(H)^J, \rho_J\} \tag{57}$$

$$\alpha_J = -\text{Tr}_J \rho_J (\tilde{S})^J + \beta_J \text{Tr}_J \rho_J (H)^J \tag{58}$$

$$\beta_J = \frac{\text{Tr}_J \rho_J \frac{1}{2}\{(\tilde{S})^J, (H)^J\} - \text{Tr}_J \rho_J (\tilde{S})_J \, \text{Tr}_J \rho_J (H)^J}{\text{Tr}_J \rho_J (H)^J (H)^J - [\text{Tr}_J \rho_J (H)^J]^2} \tag{59}$$

It is noteworthy that the functional dependence of each $D_J$ on $\rho_J$, $(\tilde{S})^J$, $(H)^J$ is the same as that of D on $\rho$, S and H for the single constituent system (Eq. 37). Proofs that Eq. (56) satisfies all the consistency conditions listed in Section 5.2 plus the three just cited are given in Refs. 12, 33, 55. Again, it is easy to verify that

$$\text{Tr}_J D_J = 0 \qquad \text{Tr}_J D_J (H)^J = 0 \qquad \text{Tr}_J D_J (\tilde{S})^J \geq 0 \tag{60}$$

and, therefore,

$$\frac{d}{dt}\text{Tr}\rho = \text{Tr}\frac{d\rho}{dt} = \text{Tr}_A D_A + \text{Tr}_B D_B = 0 \tag{61}$$

$$\frac{d}{dt}\text{Tr}\rho H = \text{Tr}\frac{d\rho}{dt}H = \text{Tr}_A D_A (H)^A + \text{Tr}_B D_B (H)^B = 0 \tag{62}$$

$$\frac{d}{dt}\text{Tr}\rho\tilde{S} = \text{Tr}_A D_A (\tilde{S}_A)(H)^A + \text{Tr}_B D_B (\tilde{S})^B \geq 0 \tag{63}$$

Finally, it is shown that if $\rho^2 < \rho$ at one time, it remains so at all times, that is for $-\infty < t < \infty$.

By taking the partial trace of $d\rho/dt$ (as given by Eq. 56) over $\mathcal{H}^B$ we obtain the rate of change of the reduced state operator of subsystem A, i.e., $d\rho_A/dt = \text{Tr}_B d\rho/dt$. If B is not interacting with A, i.e., the Hamiltonian is given by Eq. (55) with V=0, $d\rho_A/dt$ turns out to be independent $H_B$. This means that it is impossible to affect the local observables of A by acting only on B, and so nonlocality paradoxes are excluded by the novel equation of motion. This however does not mean that existing entanglement and/or correlations between A and B established by past interactions that have been subsequently turned off, have no influence whatsoever on the time evolution of the local observables of either A or B. In particular, there is no physical reason to expect that two different density operators $\rho$ and $\rho'$ such that $\rho'_A = \rho_A$ should evolve with identical local dynamics ($d\rho'_A/dt = d\rho_A/dt$) whenever A does not interact with B, because the fact that $\rho \neq \rho'$ means that in these two states the subsystems are differently correlated and/or entangled and, therefore, the two local evolutions should in general be different, at least until memory of the entanglement and the correlations established by turned-off past interactions will fade away (spontaneous decoherence) as a consequence of the irreversible entropy-increasing evolution [60]. This subtlety is captured also by the novel equation of motion. Indeed, $d\rho_A/dt$ in general depends not only on the "local" reduced density operator $\rho_A$ but also on the overall density operator $\rho$ through operator $(\tilde{S})^A = (-B\ln\rho)^A$, resulting in a collective behavior effect on the local dynamics that originates from the existing residual correlations due to past interactions.

## 6 Quantum Limits to the Second Law

### 6.1 General Remarks

The title of this section is identical to that of the First International Conference on "*Quantum Limits to the Second Law*" held in San Diego, CA in 2002 [65]. The conference was attended by 120 researchers, representing 25 countries, and 81 presentations were made. At the end of the conference an informal vote was taken on the assertion: "The second law is inviolable", and the results of the vote Yes:No:Maybe:Abstain/Absent were roughly 25:25:25:45. The Chairman of the organizing committee summarized the results by stating: "Although pains were taken to balance the conference slate, the evenness of this vote was surprising. Surely this was a select group, one not representative of the general scientific community. Were it so, we should properly consider ourselves in the midst of a fundamental paradigm shift involving the second law – which is clearly not the case".

In the light of our discussions in Sections 2-5, we affirm that the scientific community should be engaged in a revolutionary (in the sense of T. Kuhn [66]) paradigm shift in the opposite direction than that implied by the title of the conference and the results of the informal vote. The shift has occurred as a result of the recognition that entropy is an inherent – intrinsic – property of matter, represented by a unique analytical expression, and that its behavior is dictated by the equation of motion of the nonstatistical quantum theory of mechanics and thermodynamics in the same sense that, in classical mechanics, ($mv^2/2$) is an inherent property of matter, and its behavior is dictated by the equation of motion of classical mechanics.

In what follows, we review briefly many of the presentations made at and directly addressed the subject of the conference and conclude that, despite claims to the contrary, none violates the correct, unambiguous and noncircular laws and theorems of either nonstatistical quantum thermodynamics or its equivalent exposition without reference to quantum theory. For each paper that we discuss, we present our comments under a title identical to that of the paper in the proceedings of the conference.

### 6.2 Quantum Brownian motion and its conflict with the second law [67]

All claims made by the authors are false. They say:
(1) "There are not two fundamental theories of nature, quantum mechanics and thermodynamics. There is only one: quantum mechanics, while thermodynamics must emerge from it".

As we discuss earlier the situation is exactly the opposite than that claimed here. What the authors call quantum mechanics is a special case of the unified quantum theory of mechanics and thermodynamics, and corresponds to zero entropy physics.

(2) $\delta Q \leq T\delta S$ is the Clausius inequality and one of the formulations of the second law, and argue that at $T = 0$ no heat can be taken from the bath, at best heat can go from the subsystem to the bath, and conclude that the Clausius relation is thus violated at $T = 0$. In our view, both the calculations and the terminology just cited have no relation to the rigorous, well-defined, and noncircular foundations of thermodynamics, and the definition of a bath (reservoir).

(3) "The Landauer bound for information erasure is violated because the Landauer bound ... is just based on the Clausius inequality just cited." In our view, this conclusion is erroneous both because neither information erasure has any relation to the nonstatistical theory of quantum thermodynamics, nor a bath can be defined in the limit $T \to 0$.

(4) "The rate of entropy production can be negative at moderate $T$". This claim is also false because it is based on ill conceived and ill defined considerations that have no relation to thermodynamics.

(5) "Quantum Brownian motion and its conflict with the second law." This is a completely erroneous pronouncement. For a rigorous quantum thermodynamic analysis of Brownian movement see Ref. 68.

*6.3  Thomson's formulation of the second law; an exact theorem and limits of its validity [69]*

In the abstract of their presentation, the authors claim that "Thomson's formulation of the second law – no work can be extracted from a system coupled to a bath through a cyclic process – is believed to be a fundamental principle of nature. For the equilibrium situation a simple proof is presented, valid for macroscopic sources of work. Thomson's formulation gets limited when the source of work is mesoscopic, i.e., when its number of degrees of freedom is large but finite. Here work-extraction from a single equilibrium thermal bath is possible when its temperature is large enough. This result is illustrated by means of exactly solvable models. Finally we consider the Clausius principle: heat goes from high to low temperature. A theorem and some simple consequences for this statement are pointed out."

These claims are false. By definition, a "bath" or "reservoir" is a system from which any amount of energy $Q$ extracted is unavoidably accompanied by an amount of entropy $Q/T_R$, where $T_R$ is the constant temperature of the reservoir, and $Q$ is independent of $T_R$. It follows that no work (energy only) can be extracted from $R$ by a cyclic process, because work is an interaction defined as an exchange of energy only.

For completeness, we emphasize that energy flows from a system at negative temperature to a system at positive temperature but then the latter is not a reservoir [10].

*6.4  The second law and the extension of quantum mechanics [70]*

In the introduction of their paper, the authors aver: "Physicists are used to consider as fundamental two theories, namely thermodynamics at the macroscopic level and quantum theory at the microscopic level. Up to now, all experimental features are in accordance with both approaches. None of those theories has found a limitation in its application. Nevertheless, those theories do not share the same symmetry properties with respect to time. On the one hand, the usual quantum theory provides a time reversal invariant description while on the other hand, the second law of thermodynamics emphasizes the irreversible nature of the physical world. Their compatibility can therefore be questioned, at least at the conceptual level. A domain where their incompatibility has generated many philosophical turmoil is in circumstances where both theories are required simultaneously. Such a case has been prevalent from the early days of quantum theory for the understanding of its interpretation. The measurement act in quantum physics requires a macroscopic apparatus and irreversibility to freeze its results and make the choice among the possible outcomes. Indeed, as long as no irreversibility occurs in the system, no reading of an instrument can be performed since the measurement process could be undone, at least at the conceptual level (the physical system is still described by a state vector which is a combination of vectors leading to all possible outcomes). That problematics has led to many oddities such as the Schrödinger cat, the Wigner's friend, the many-worlds interpretation of Everett, the mind-matter debate, ... Certain eminent physicists solve the problem by simply denying it (For Einstein, irreversibility is a human illusion). Others want to prove that such difficulties are irrelevant in practice due to a decoherence time out of any foreseeing observation. Such an approach leaves open the conceptual debate, the large audience at this meeting gives evidence that many people are aware of the problems involved."

It is absolutely correct that statistical theories of thermodynamics yield many accurate and practical numerical results about stable – thermodynamic – equilibrium states. As we discuss earlier, however, the almost universal efforts to compel thermodynamics to conform to statistical explanations is not justified in the light of many equally accurate, logically consistent, and reproducible nonstatistical experiences that resulted in dramatic changes of the scientific paradigm. Accuracy of predictions is a necessary requirement of any theory of physics but is not sufficient if the predictions are incomplete. Outstanding examples of the importance of completeness are the ideas of Aristarhus and Galileo Galilei that contributed to our understanding that the solar system is heliocentric and not geocentric without affecting numerical results.

The comment that "macroscopic thermodynamics and quantum theory do not share the same symmetry properties with respect to time" is not correct for several reasons. As we discuss earlier, thermodynamics does not deny the existence of conventional quantum mechanics and classical mechanics (zero entropy physics) which use the same concept of time as that in the unified quantum theory of mechanics and thermodynamics. In fact, the laws of thermodynamics in general, and the characteristics of entropy, in particular, do not require an asymmetry in time. The reasons are the following two important theorems of the laws of thermodynamics: (i) If an adiabatic process is reversible, the entropy remains invariant; and (ii) If an adiabatic process is irreversible, the entropy increases. The key word in these two theorems is "if". In other words, the laws of thermodynamics do not require that processes be reversible or irreversible. These laws simply predict the consequences of each of these two types of processes.

In addition to theoretical considerations, the conclusion just cited is of paramount practical importance. If processes were required to be irreversible, then we would be faced with the dilemma "what is the lowest limit of irreversibility?" Fortunately, no one has a nonzero answer to this question, and over the past two and a half centuries, engineers are designing devices that approach the zero limit, even though they started with more than 99 percent losses attributable to irreversibility two and a half centuries ago. So there exists no fundamental incompatibility between quantum mechanics and thermodynamics.

The association of irreversibility with quantum measurements is unwarranted, and misrepresents the foundations of quantum theory (see Section 4).

Schrödinger's cat does not represent a paradox because no projector or density operator can be expanded into (let alone be considered as a superposition of) two terms, one valid at a time $t_1$ and the other at time $t_1 + \tau$. All expansions (not superpositions) in quantum theory consist of terms defined at one instant in time. Besides in the case of Schrödinger, the systems radioactive source and live cat are radically different before and after the death of the cat, and the decay of the radioactive source.

Moreover, Einstein did not ever say that "irreversibility is a human illusion". What he said was: "For us physicists this separation between past, present, and future holds only an illusion, tenacious as it may be [71]". In view of Einstein's remarks, and the observation that the laws of physics (either Thermodynamics or Quantum Thermodynamics) do not dictate that phenomena must be irreversible, we conclude that there is no arrow of time. Time is a dimension along which phenomena evolve either forward or backward.

We feel that questions of coherence and decoherence arise from lack of clarity in definitions of fundamental concepts such as system, property, state, measurement results, and expectation values (see Section 4).

We do not comment on the theory presented at the conference by deHaan and George [70] because the authors state that "it is not finished". However, we would like to express our full agreement with the summary of their position. To wit, they say: "Summarizing, at least one of the two fundamental theories has to be modified.[13] The title of the conference seems to induce that quantum world has to impose its view on thermodynamics and would restrict the impact of the second law. Our view is that the opposite attitude has to prevail (obviously, in the strict respect of all the achievements of quantum physics). If thermodynamics has to be kept entirely, without quantum limits, it means that the quantum theory has to be enlarged to encompass new situations".

We could not agree more with the view just stated, and feel that our small group at MIT has made significant progress in the direction recommended by Professors de Haan and George as evidenced by our discussions in Sections 4 and 5.

### 6.5 *Extracting work from a single bath via vanishing quantum coherence [72]*

In part of the abstract, the author says: "We here extend the previous macroscopic (quantum thermodynamical) analysis by developing a microscopic (quantum statistical) analysis. This provides insight into engine operation and questions related to the second law. The same quantum phase effects that yield lasing without inversion and ultraslow light, also make possible extensions of Carnot cycle operation; e.g., extraction of energy from a single heat bath, and efficiency beyond the Carnot limit."

We have very serious and fundamental objections to the statements made in the abstract and the related arguments in the text of this presentation at the Conference for the following reasons:

(1) The correct terminology for a heat bath is simply *reservoir* because any system in any state (thermodynamic equilibrium or not thermodynamic equilibrium) contains neither heat nor work. Heat and work are not properties of a system. They are modes of interactions. By definition, a reservoir is a system that experiences only heat interactions with other systems, that is, interactions that involve the exchange of both energy $Q$ and entropy $Q/T_R$, where $T_R$ is the constant temperature of the reservoir for all values of energy $Q$, and $Q$ is independent of $T_R$. But these characteristics do not mean that a reservoir contains heat. It follows that no work can be extracted from a reservoir because work is an interaction that involves the exchange of energy $W$ only. If work could be extracted from a reservoir, then that would have been a realization of a Maxwell demon, a realization that, for the first time, has been recently shown to be impossible for any system in a thermodynamic equilibrium state [73].

(2) The *thermodynamic efficiency* of a Carnot cycle experiencing only reversible processes is 100%. It is the ratio of the work output over the availability (Keenan's definition) or exergy (terminology used in many textbooks but synonymous to Keenan's availability). So one cannot have a thermodynamic efficiency greater than 100%. To be sure, the same engine, as the one just cited, has a thermal efficiency (ratio of work output to heat input) less than 100% but, whatever the value, it cannot be improved because the thermodynamic efficiency would be greater than 100%.

---
[13] The two theories are thermodynamics and quantum physics.

(3) Now, to increase the work output of the preceding engine one must utilize additional energy and, therefore, availability than that employed by the perfect engine just discussed. In particular, the author of this contribution reports that the amount required is 1010 units in order to create 1 unit of work in the form of "phaseonium" [72]. Even if all the processes associated with the phaseonium utilization were thermodynamically perfect, the efficiency of utilization (work out over energy in) would be less than 0.1%, approximately what steam engines could achieve two and a half centuries ago. So such a process, even if feasible, would be a great waste of energy resources.

*6.6 Dimer as a challenge to the second law [74]*

In the abstract of this presentation, the author asserts that: "Even such a simple system as an asymmetric dimer cooperating with two baths can be used to challenge the Second law of thermodynamics. Only a specific coupling to the baths and a specific regime is needed". However, this assertion is not definitively supported by the ambiguous conclusions of the text.

In his introduction, the author cites ten references and declares that: "In view of experimental, theoretical, as well as combined evidence, it is now hardly answerable to simply disregard or ignore *a priori* all challenges to the Second law of thermodynamics as erroneous, not reproducible, or misleading. One should add that a number of other paradoxical systems exist where experimental evidence speaks in favour of the Second law violations though decisive experiments have not been performed so far. Growing interest in the problem that would either deepen our understanding of Nature or potentially blow up all the thermodynamics as a universal scientific discipline implies also increasing interest in, and importance of looking for, as simple systems as possible that would not mask essence of the relevant processes by technical complexity. Single quantum Brownian particle in a harmonic confining potential and coupled linearly to a bath of non-interacting harmonic oscillators is perhaps still too simple for such purposes. Here, we report, however, results of our analysis showing that already a simple quantum particle on a pair of non-equal states, i.e., an asymmetric dimer, may under favourable conditions behave in a way fully challenging the Second law. These conditions include a specific type of coupling to a pair of baths as well as the regime of a sufficiently strong coupling of this type."

On the basis of his analysis, the author concludes "... spontaneous heat transfer from our colder bath I to the warmer bath II. Both the baths are, however, owing to the thermodynamic limit preceding the long-term one, macroscopic. This should make standard thermodynamics applicable. Such a spontaneous heat flow from colder to warmer macroscopic bodies as obtained above should, however, never appear according to the Clausius formulation of the Second law of the phenomenological thermodynamics". But then, he contradicts himself by saying "... that though the model and its treatment are rigorous, arguments as above are still open to question .... Hence verification of the above conclusions by independent methods is needed".

In our view, this kind of discussion does not present a challenge to the laws of thermodynamics. We say the laws of thermodynamics because, as we discuss earlier, the complete definition of entropy, temperature, and heat require all three laws.

## 6.7 Macroscopic potential gradients: Experimentally-testable challenges to the second law [75]

In the abstract of his presentation, the author claims that: "Many equilibrium systems exhibit potential gradients (e.g., gravitational, electrostatic, chemical). Over the last ten years these have been implicated in a series of second law challenges spanning gravitational, chemical, plasma, and solid state physics. Here they are reviewed with an eye toward experimental tests. Macroscopic potential gradients are advanced as a new predictor of second law challenges".

In view of the fundamental principles and theorems of thermodynamics, and of the nonstatistical quantum thermodynamics discussed earlier, we find that none of the claims in this paper represents a predictor of a second law challenge. Specifically, the *gravitational challenge* is described as consisting of "a tenuous gas contained in a planetary-sized cavity that houses a moon-sized spherical mass which is paved asymmetrically; one hemisphere is elastic, while the other is inelastic with respect to suprathermal gas collisions. Gas cycling through the gravitational potential gradient between the gravitator and cavity walls exerts a new unbalanced force on the gravitator by which it can move and produce steady-state work (e.g., turn an electrical generator). Three-dimensional numerical simulations of test atoms in the cavity verify that over a wide range of realistic physical conditions, power should be extractable from such a system. Laboratory experiments involving suprathermal argon and helium beams colliding with different surface types verify the pivotal asymmetry in gas-surface collisions upon which this paradox depends."

This example represents neither a paradox, nor a challenge to thermodynamics for the following reasons: (i) By definition, a blackbody is in a stable – thermodynamic – equilibrium state and, therefore, the photons are not flowing, and cannot impart momentum to the pliable sphere; and (ii) If and only if the pliable sphere is not in a state in mutual stable equilibrium with the radiation, then and only then the composite system can do work on its environment, and such work is not a challenge to the laws of thermodynamics.

A simple example of the preceding discussion is a wrist watch and its tiny battery. Because the tiny battery is not in mutual stable equilibrium with the watch, and has a very long time constant of spontaneous internal discharge, the watch works very well over many years.

Next, the author considers the case of Plasma I, and describes it as follows: "An electrically conducting probe is suspended in a high-temperature, plasma-containing blackbody cavity that is coupled to a heat bath. The cavity walls are thermally coupled to the heat bath and the probe is connected to the cavity walls through a load via a switch. When the probe is physically disconnected from the walls (ground), it electrically charges as a capacitor to the plasma floating potential. When the switch is closed, the probe discharges as a capacitor through the load and plasma. With an ideal switch, this charging and discharging of the probe through the load can be repeated indefinitely. Laboratory experiments indicate a non-zero voltage and current from a small probe can persist under blackbody cavity plasma conditions and can be varied by adjusting the plasma density, in agreement with theoretical predictions".

The analysis of this experimental situation is not correct. The capacitor will be charged if and only if it is not in mutual stable equilibrium with the plasma. Upon reaching mutual stable equilibrium, no work can be done because that would be a PMM2, a machine that violates both the laws and theorems of the thermodynamics, and the laws and theorems of nonstatistical quantum thermodynamics. So this situation is not a challenge to the laws of thermodynamics.

Next, the author considers the case of Plasma II, and describes it as follows: "This system consists of a frictionless, two-sided piston in a high-temperature, plasma-filled blackbody cavity surrounded by a heat bath. Opposing piston faces are surfaced with different work function materials (surfaces S1 and S2). Owing to differential neutral, electronic and ionic emissions from the different surfaces, a steady-state pressure difference is sustained between the piston faces, which can be exploited to do work. Laboratory experiments testing a critical aspect of the paradox – that different surfaces can simultaneously thermionically emit distinctly in a steady-state fashion in a single blackbody cavity – corroborate theoretical predictions".

This example also misrepresents the laws and theorems of thermodynamics. For a complete and rigorous analysis that proves that no current flows either out of or into any of the surfaces under any of the specified conditions see [76, 77]. So no challenge emerges from the plasma examples.

Finally, we reach similar conclusions about the so-called chemical and solid state challenges because they too are based on misinterpretations and misconceptions about thermodynamics.

*6.8 Measuring the temperature distribution in gas columns [78]*

In the abstract of his presentation the author states that: "Late in the 19th century J. Loschmidt believed that a vertical column of gas in an isolated system would show a temperature gradient under the influence of gravity, cold at the top and warm at the bottom. L. Boltzmann and J.C. Maxwell disagreed. Their theories tried to prove an equal temperature over height. Experiments with various test setups are being presented which seem to strengthen the position of Loschmidt. Long term measurements at room temperature show average temperature gradients of up to 0.07 K/m in the walls of the enclosure, cold at the top and warm at the bottom. The measured values can be explained by the conversion of the potential energy of the molecules into an increase of their average speed through gravity".

Furthermore, in the concluding summary, the author states: "Measuring the temperature distribution in isolated spaces filled with a gas and a powder a vertical temperature gradient was found, cold at the top and warm at the bottom as argued by J. Loschmidt. This result seems to be a contradiction to the 2nd law of thermodynamics. If correct, it would make possible the creation of work out of a heat bath".

Unfortunately for the energy hungry humanity, the concluding remark is not valid. The issue examined by the author is rigorously analyzed by Hatsopoulos and Keenan [14] who find that in a gravity field the total and chemical potentials $\mu_i$ and $\mu_{ci}$ of the $i$-th constituent of a gas column in a stable equilibrium state satisfy the relation

$$\mu_i = \mu_{ci} + \gamma \qquad (64)$$

where $\gamma$ is such that grad $\gamma = -g$, and $g$ is the intensity of the gravity field. Moreover, by definition a heat bath is isothermal (grad T=0), and has no gradients of total potentials (grad $\mu_i = 0$ for all $i$). So there is no contradiction with the laws of thermodynamics.

*6.9 Violation of the third law of thermodynamics through an atomic cooling scheme [79]*

The author correctly quotes a thermodynamic opinion that expresses the third law in the form "It is impossible by any procedure, no matter how idealized, to reduce the temperature of any system to absolute zero in a finite number of operations," and comments that "This is a correct statement under the usual macroscopic schemes due to the fact that heat capacity vanishes as one approaches absolute zero temperature. However, the scheme we present here is not so limited. Our observation is that the third law need not apply to the process of cooling by coherent control. Thus, in fact, it is possible to cool to absolute zero in a rather small number of steps. Consequently, the third law is not an entirely general law".

As we discuss earlier, the correct thermodynamic statement of the third law is: "*for each given set of values of the amounts of constituents and the parameters, there exists one stable* *equilibrium state with zero temperature* (if the system has no upper bound on energy), and *two stable equilibrium states with zero temperature* (if the system has an upper bound on energy, such as a one spin system). Therefore, there are no two third laws, one for macroscopic schemes and another for microscopic processes. Besides, the correct three laws of thermodynamics are theorems of the unified, nonstatistical quantum theory of mechanics and thermodynamics.

More generally, however, were we to deny the two extreme temperatures of $1/T = \infty$ and $-\infty$ then we would be denying the existence of the two (up and down) orientations of a spin, and the existence of a ground state of a system without upper limit on energy ($1/T = \infty$). None of these denials is part of thermodynamics.

*6.10 Perpetual motion with Maxwell's demon [80]*

The author claims that: "A method for producing a temperature gradient by Brownian motion in an equilibrated isolated system composed of two fluid compartments and a separating adiabatic membrane is discussed. This method requires globular protein molecules, partially embedded in the membrane, to alternate between two conformations which lie on opposite sides of the membrane. The greater part of each conformer is bathed by one of the fluids and rotates in Brownian motion around its axis, perpendicular to the membrane. Rotational energy is transferred through the membrane during conformational changes. Angular momentum is conserved during transitions. The energy flow becomes asymmetrical when the conformational changes of the protein are sterically hindered by two of its side-chains, the positions of which are affected by the angular velocity of the rotor. The heat flow increases the temperature gradient in contravention of the Second Law. A second hypothetical model which illustrates solute transfer at variance with the Second Law is also discussed".

The proposed methods do not and cannot violate either the laws of thermodynamics or the foundations of quantum thermodynamics. Brownian motion is observed in systems that consist of a solvent and a colloid each of which experiences no change in energy, amounts of constituents, volume, and entropy and therefore no change in temperature, chemical potentials, and pressure. What is observed is not caused by motions but by infinitely large differences between total potentials of constituents of the colloid and the same constituents that are absent from the solvent, and conversely total potentials of constituents of the solvent and the same constituents that are absent from the colloid [68]. It is these differences that affect the pliable shape of the interface between the solvent and the colloid and appear to an observer as motions, and not motions of the constituents as such that bring about what appears as motions. It is clear that the systems analyzed in Ref. 80 do not consist of a solvent and a colloid, and do not violate any laws of thermodynamics. In addition, the systems contemplated in Ref. 80 have no relation to the systems on which Maxwell's brain child is supposed to perform his diabolical acts as we will see in Section 6.13.

*6.11 Bath generated work extraction in two level systems [81]*

In the abstract of this presentation the authors declare: "The spin-boson model, often used in NMR and ESR physics, quantum optics and spintronics, is considered in a solvable limit to model a spin one-half particle interacting with a bosonic thermal bath. By applying external pulses to a non-equilibrium initial state of the spin, work can be extracted from the thermalized bath. It occurs on the timescale $\mathcal{J}_2$ inherent to transversal ('quantum') fluctuations. The work (partly) arises from heat given off by the surrounding bath, while the spin entropy remains constant during a pulse. This presents a violation of the Clausius inequality and the Thomson formulation of the second law (cycles cost work) for the two-level system."

The analysis and conclusions are not correct because the authors fail to recognize the definition of a thermal bath or reservoir. Ever since the path blazing discussions of Carnot, and the pioneering contributions of Clausius, a bath or reservoir is defined as a system that for every value of its energy is in a stable (thermodynamic) equilibrium state, and that all these states have the same temperature. Accordingly, each amount $Q$ of energy received or delivered in the course of an interaction of a reservoir with any system is accompanied by an amount of entropy received or delivered equal to $Q/T_R$, where $T_R$ is the fixed temperature of the reservoir, and $Q$ is independent of $T_R$. An interaction that involves energy $Q$ and entropy $Q/T_R$ is called heat.

In view of these well defined concepts, it is clear that no work can be extracted from a reservoir because a work interaction, by definition, involves only energy exchange.

### 6.12 *The adiabatic piston and the second law of thermodynamics* [82]

In the abstract of this presentation, the authors claim that: "A detailed analysis of the adiabatic-piston problem reveals peculiar dynamical features that challenge the general belief that isolated systems necessarily reach a static equilibrium state. In particular, the fact that the piston behaves like a *perpetuum mobile*, i.e., it never stops but keeps wandering, undergoing sizable oscillations, around the position corresponding to maximum entropy, has remarkable implications on the entropy variations of the system and on the validity of the second law when dealing with systems of mesoscopic dimensions".

The interpretation of this problem is not correct because there are no "perpetuum mobile", and no "remarkable implications" ... on the validity of the second law when dealing either with systems of mesoscopic dimensions (or any dimensions, beginning with just one spin). At the end of the process, the two systems on the two sides of the piston are in *partial mutual stable equilibrium* with respect to temperature, that is, the temperature of one system differs from the temperature of the other. To see this result clearly, we recognize that the problem has five unknown quantities to be evaluated, i.e., two final temperatures, two final volumes, and one common final pressure. The five unknown values just cited satisfy five equations, i.e., two energy equations in terms of temperatures and pressures, one total energy conservation equation, one total volume conservation equation, and one equality of final pressures. The reason for the misunderstanding of the problem by Callen is that he tries to make his calculations by assuming that the two interacting parts are passing only through stable equilibrium states. This assumption is consistent with Callen's conception of thermodynamics (see Section 3.5) but wrong for the general theory of thermodynamics summarized in Sections 2 and 3.

Phenomena of partial mutual stable equilibrium arise frequently in thermodynamics. One example is the experimental procedure for the measurement of partial pressures of a mixture by means of rigid membranes, each permeable by one constituent only. Across each of these membranes, two of the requirements for mutual stable equilibrium are satisfied, i.e., temperature and total potential equalities, but not pressure equality, i.e, the pressure of the mixture is not equal to the so called partial pressure of the one constituent system in partial mutual stable equilibrium with the mixture.

### 6.13 *Maxwell's demon* [83–86]

Several presentations address the question of the feasibility of Maxwell's demon. We are not going to comment on each presentation separately because all claim to exorcise the demon but in our view none confronts the problem posed by Maxwell.

In concert with every other discussion of Maxwell's demon that has ever been published (more than 400 papers in the archival literature, and many books), the presentations at the con-ference (see however Section 6.14) misrepresent and misinterpret the specifications of the sharp-witted being conceived by Maxwell, and later named a demon by Thomson. Maxwell defined his omnipotent and omniscient brain child as an entity that can do anything he pleases at no cost whatsoever to himself or any physical system, be it in the form of energy, entropy, information, erasure of information, ratchet and pawl performance, or a new invention of specific gadgetry or anything imaginable. One may think that such a boundless specification is unrealistic. But whether we feel that he disregarded the laws of physics, and whether we like it or not that is Maxwell's conception of his demon.

Of the myriad of publications on the subject, only two address and resolve the problem specified by Maxwell. One is purely thermodynamic [73]. We show that despite his omnipotence and omniscience, the demon cannot accomplish his task, namely extract only energy from the system or, equivalently, have the system do work at no energy (work) cost whatsoever to himself, because under the specified conditions there exist no states with energy lower than the initial energy of the system. Said differently, there exists no state that has identical values of all the properties as the initial thermodynamic equilibrium state except lower energy.

The nonexistence of states of lower energy that satisfy the specifications of the problem is not the result of a tautological and/or circular argument such as the Kelvin-Planck statement of the second law, or the Caratheodory statement of the second law. It is the result of theorems of thermodynamics that differs radically from those in the literature of physics and thermodynamics. It is the result of the exposition of thermodynamics in Sections 2 and 3.

The second proof of the impossibility of the demon to violate the laws of physics is quantum thermodynamic [87] without statistical probabilities (see Sections 4 and 5). We show that the demon cannot accomplish his task starting from a thermodynamic equilibrium state because in any such state there are no moving molecules, each molecule has a zero velocity. Accordingly, no matter how omnipotent and omniscient the demon is, there are no fast and slow molecules to be sorted out according to Maxwell's specifications. We realize that upon reading this conclusion of quantum thermodynamics, some of our readers will throw our manuscript to the waste-paper basket, and yet our conclusion is rigorously proven.

Some other readers may agree that the expectation value of the velocity $\langle p_i \rangle / m_i = v_i = 0$, where $p_i$ is the momentum operator, and $v_i$ and $m_i$ the velocity and mass are the velocity and mass of the $i$-th molecule, but then would counterargue that $\langle p_i^2 \rangle / 2m_i \neq 0$ and therefore each molecule has kinetic energy. But that would be a misinterpretation of the rules of quantum theory about when it is possible to interpret a quantum expression by its classical equivalent [36] because we would have to write $\langle p_i^2 \rangle / 2m_i = \langle p_i \rangle^2 / 2m_i = m_i v_i^2 / 2$ thus concluding that $\langle p_i^2 \rangle - \langle p_i \rangle^2 = (\Delta p_i)^2 = 0$ instead of $\Delta p_i \neq 0$. The impossibility of using the classical expression to describe kinetic energy is not unique to quantum theory. For example, in the general theory of relativity there is only energy, and no distinctions can be made about different forms of energy.

### 6.14 Who's afraid of Maxwell's demon – and which one? [88]

In the abstract of his presentation, the author, Professor Callender, states: "Beginning with Popper, philosophers have found the literature surrounding Maxwell's demon deeply problematic. This paper explains why, summarizing various philosophical complaints and adding to them. The first part of the paper critically evaluates attempts to exorcise Maxwell's demon; the second part raises foundational questions about some of the putative demons to be summoned at this conference."

In our view, this is an excellent discussion of the issues raised by the multitude of publications on Maxwell's demon, as evidenced by the author's opening statement: "Beginning with Popper, philosophers examining the literature on Maxwell's demon are typically surprised – even horrified. As a philosopher speaking at a physics conference exactly 100 yrs after Popper's birth, I want to explain why this is so. The organizers of this conference instructed me to offend everyone, believers and non-believers in demons. Thus my talk, apart from an agnostic middle section contains a section offending those who believe they have exorcised the demon and a second offending those who summon demons. Throughout the central idea will be to clearly distinguish the various second laws and the various demons. Since to every demon there is a relevant second law, and vice versa, my talk will therefore be a kind of map of the logical geography of the underworld of physics."

We are happy to report that our publications on the subject [73, 87] do not suffer by any of the maladies discussed by Professor Callender, because we have provided two definitive answers which would not horrify the late Philosopher Popper.

The only cautionary note we wish to add is about Professor Callender's discussion of "Nonequilibrium steady state Maxwell's demons". There is no problem. Energy and nothing else can be extracted from a state without violating the laws of thermodynamics (i.e. a work interaction – adiabatic process – is possible) if the state of the system is not a stable equilibrium state, that is, if the state is unstable equilibrium, nonequilibrum, steady or unsteady because all such states have an adiabatic availability [see Section 2.7] greater than zero. The limitation on energy extraction – work interaction – exists only for thermodynamic equilibrium states that correspond to fixed values of energy, amounts of constituents, and volume [71, 86].

## 7 Challenges to the Second Law of Thermodynamics

The title of this section is identical to that of a monograph by Čápek and Sheehan [89], sequel to the presentations made at the conference discussed in Section 6. In what follows, we comment on some remarks made in the monograph, and conclude that they represent neither a clear understanding of the issues raised by the usual expositions of thermodynamics, nor any challenges to the second law of thermodynamics.

For example, on p. xiii the authors declare that: "The second law has no general theoretical proof and, like all physical laws, its status is ultimately tied to experiment. Although many theoretical challenges to it have been advanced and several corroborative experiments have been conducted, no experimental violation has been claimed and confirmed. It is our position, however, given the strong evidence for its potential violability, that inquiry into its status should not be shifted by certain unscientific attitudes and practices that have operated thus far". But if despite the strong evidence for the violability of the second law no violation has been confirmed, what is the rationale for characterizing those that are convinced about the validity and importance of the correct exposition of thermodynamics as individuals that represent unscientific attitudes and practices?

Next, on p. xiv, the authors conclude their remarks by asserting that: "It is likely that many of the challenges in this book will fall short of their marks, but such is the nature of exploratory research, particularly when the quarry is as formidable as the second law. It has 180 years of historical inertia behind it and the adamantine support of the scientific community". If no challenges have been proven valid, what is the motivation for pursuing exploratory research to prove that the second law is invalid? To put our question differently, why people interested in exploratory research do not try to prove that the solar system is neither geocentric nor heliocentric? Similarly why researchers do not try to prove that, in the realm of its validity, Newton's equation of motion is not correct?

In Chapter 1 of their book, Čápek and Sheehan summarize twenty-one formulations of the second law, and twenty one varieties of entropies. Their discussions, however, reveal complete misunderstandings of the issues that surround the foundations of the science of thermodynamics, the acceptability of thermodynamics, and the acceptability of the proposed resolutions. In particular, they completely misunderstand and misrepresent the work that we have summarized in Sections 2-5. Among many other examples, a simple illustration of the judgement just cited is given by remarks in pp. 23-24 of the book. The authors state: "There is no completely satisfactory definition of entropy. To some degree every definition is predicated on physical ignorance of the system it describes and, therefore, must rely on powerful *ad hoc* assumptions to close the explanatory gap. ... Finally, as it is deduced from thermodynamics, the notion of entropy is critically dependent on the presumed validity of the second law. ... Some entropies, like the Gyftopoulos, Hatsopoulos, Beretta entropy are claimed to apply at nonequilibrium, but they do not have compelling microscopic descriptions".

Even a cursory study of our work reveals that none of the preceding statements are valid. Our definition of entropy is not predicated on physical ignorance of the system, and does not rely on *ad hoc* assumptions. The thermodynamic definition of entropy is a theorem of both the first and the second laws, but neither the first nor the second law contains the concept of entropy. Moreover, in the quantum thermodynamic exposition, all the laws of thermodynamics are theorems of the general equation of motion. Finally, the concept of entropy that we derive applies to nonequilibrium states as well as to any other kind of state, and has "compelling microscopic descriptions" because it is shown to be a measure of the quantum mechanical spatial shape of the constituents of the system in any state.

Other examples that support our earlier judgement are the discussions of conflation between entropy and the second law, and the introduction of the zeroth and third laws of thermodynamics in pp. 26-29 of the book. Some peculiar statements are as follows:

(1) "At the microscopic level an individual molecule doesn't know what entropy is and it couldn't care less about the second law". This is an unjustified, and completely unacceptable remark. All one has to consider is the effects of human activities on natural resources and our environment. They are real and they are entirely due to the increase of entropy. So both individual molecules and living beings on planet earth must care about the laws of thermodynamics.

(2) "In its very conception, entropy presumes ignorance of the microscopic details of the system it attempts to describe". This pronouncement is not correct. As an intrinsic property of each constituent of the theories of physics, entropy knows all the details, and has a definite impact on the effectiveness of the constituents it affects as they evolve in time.

(3) In the opinion of one of the authors (d.p.s): "... entropy will never have a completely satisfactory and general definition, nor will its sovereign status necessarily endure ... Entropy remains enigmatic. The more one studies it, the less clear it becomes ..." These statements are not only false in the light of our discussions in Sections 2-5, but they are also nonscientific in view of the lack of even a nanotrace of supportive evidence.

(4) *Zeroth law*. The authors cite the zeroth law in terms of the following statement: "If the temperature of system *A* is equal to the temperature of system *B*, and the temperature of system *B* is equal to the temperature of system *C*, then the temperature of system *A* is equal to the temperature of system *C*". A zeroth law is neither required nor used in the exposition of thermodynamics. The temperature equalities just cited emerge from the fundamental relation [10].

(5) *Third law*. Nernst-Planck: "Any change in condensed matter is, in the limit of the zero absolute temperature, performed without change in entropy". This is not correct. Beginning from a state at zero temperature, changes are possible with $\Delta S \geq 0$, where $\Delta S = 0$ applies if the change occurs from the projector that corresponds to the ground state to another projector that corresponds to an energy larger than the energy of the ground state, and $\Delta S > 0$ if the final state is described by a density operator $\rho > \rho^2$. Here, the entropy change is accompanied by an energy change, and the largest entropy increase – entropy exchanged plus entropy generated by irreversibility – is delimited by the fundamental relation of the condensed matter.

(6) *Third Law*. Planck: "The entropy of any pure substance at $T = 0$ is finite and, therefore, can be taken to be zero". This is not correct. The entropy of any substance $S(T = 0) = 0$ because the density operator is a projector, and all projectors have $S$(projector)=0.

(7) Next Čápek and Sheehan introduce their own statements of the third law, to wit: (a) It is impossible to reduce the temperature of a system to absolute zero via any finite sequence of steps; and (b) *Perpetuum mobile* of the third types are impossible. Both statements are not correct. Statement (a) is invalid because if it were correct it would imply that thermodynamics denies the existence of ground states of different systems, an implication that is false; and statement (b) is not correct because the third law has no relation to *perpetuum mobile*.

(8) The authors close their remarks in Chapter 1 by the following statements: "In summary, the laws of thermodynamics are not as sacrosanct as one might hope. The third law has been violated experimentally (in at least one form); the zeroth law has a warrant out for its arrest; and the first law can't be violated because it's effectively tautological. The second law is intact (for now), but as we will discuss, it is under heavy attack both experimentally and theoretically". Despite the interesting sense of humor, the preceding closing remarks are baseless, nonscientific, and have nothing to do with the beautiful, powerful, and faultless general theory of thermodynamics without quantum theoretical considerations (Sections 2 and 3), and the unified quantum theory of mechanics and thermodynamics without statistical probabilities (Sections 4 and 5).

Besides, despite their good intentions, Čápek and Sheehan have already declared that "Although many theoretical challenges to the second law have been advanced and several corroborative experiments have been conducted, no experimental violation has been claimed and confirmed". Accordingly, we are not going to discuss the remaining chapters of their book.

## 8 Conclusions

To facilitate the answer to the question of the validity of the second law of thermodynamics, we provide extensive summaries of two novel but intimately interrelated expositions of thermodynamics, one without explicit reference to quantum theory, and the second quantum mechanical without statistical probabilities of the kinds used in either statistical classical mechanics or statistical quantum mechanics.

We use the definitions, postulates, and theorems of the two expositions just cited, and evaluate the correctness of claims of violations of the "second law of thermodynamics". We find that none of the claims is justified, and none represents a limit to the validity of the unambiguous and/or noncircular statement of the second law of thermodynamics. In fact, the new expositions of the science of thermodynamics provide the tools for extension of applications to states that are not thermodynamic equilibrium and a more general regularization of natural phenomena than existed up to now. As such, they hold the promise of better scientific and engineering applications, and of more effective use of natural resources.